\documentclass[enabledeprecatedfontcommands,11pt,a4paper]{scrartcl}

\usepackage[utf8]{inputenc}
\usepackage[T1]{fontenc}
\usepackage[english]{babel}
\usepackage[autostyle]{csquotes}

\usepackage{lmodern}
\usepackage[pdftex]{graphicx}
\usepackage{pdfpages}

\usepackage[noblocks]{authblk}
\usepackage{orcidlink}
\usepackage[nointegrals]{wasysym}
\usepackage{fontawesome5}
\usepackage{sty/repostatus}

\usepackage[all]{nowidow}
\usepackage{multicol}
\usepackage{amsmath} 
\usepackage{booktabs} 
\usepackage{threeparttable} 
\usepackage{hologo} 

\usepackage[margin=1.5cm, includefoot]{geometry}
\usepackage{fancyhdr}
\usepackage{enumitem}
\setitemize[itemize]{itemsep=0.1em, parsep=0em, topsep=0.2em}
\setitemize[enumerate]{itemsep=0.1em, parsep=0em, topsep=0.2em}

\usepackage{tcolorbox} 
\usepackage[colorinlistoftodos,prependcaption]{todonotes} 

\usepackage[style=numeric,datamodel=software]{biblatex}
\usepackage{software-biblatex}
\usepackage[toc,section=section]{glossaries}

\usepackage{hyperref}
\hypersetup{
    pdfauthor={Druskat, S.; Bertuch, O.; Juckeland, G.; Knodel, O.; Schlauch, T.},
    pdfkeywords={Research software; Software publication; Software metadata; Automation; Deposition; Workflows; Repositories; Metadata extraction; Metadata harvesting; Metadata formats; FAIR; FAIR4RS; Provenance; Reproducibility},
    pdflang={EN},
    backref,
    bookmarksopen=true,
    breaklinks=true,
    colorlinks=true,
    linkcolor=blue,
    citecolor=blue
}

\usepackage{caption}
\usepackage{bookmark}
\makeatletter
    \pretocmd\endtable{%
      \bookmark[
        rellevel=1,
        keeplevel,
        dest=\@currentHref,
      ]{Table \thetable: \@currentlabelname}%
    }{}{\errmessage{Patching \noexpand\endtable failed}}
    \pretocmd\endfigure{%
      \bookmark[
        rellevel=1,
        keeplevel,
        dest=\@currentHref,
      ]{Figure \thefigure: \@currentlabelname}%
    }{}{\errmessage{Patching \noexpand\endfigure failed}}
    
    \AtBeginDocument{
        \lfoot{\enquote{\@title} \- v\versionnumber}
        
        \hypersetup{
            pdftitle={\@title \- v\versionnumber},
            pdfsubject={\@subtitle: current state of metadata for software and a concept to enhance research software publications with them in an automated fashion}
        }
    }
\makeatother

\usepackage{crossreftools} 
\usepackage{subfiles} 

\newcommand*\versionnumber{1}
\newcommand*\version{\\ {\normalsize Version \versionnumber{}}}
\newcommand{\fn}[1]{\texttt{#1}}

\title{Software publications with rich metadata}
\subtitle{State of the art, automated workflows and HERMES concept}

\date{\today \version}

\author[1,*]{Druskat, Stephan \orcidlink{0000-0003-4925-7248}}
\author[2]{Bertuch, Oliver \orcidlink{0000-0002-2702-3419}}
\author[3]{Juckeland, Guido \orcidlink{0000-0002-9935-4428}}
\author[3]{\authorcr Knodel, Oliver \orcidlink{0000-0001-8174-7795}}
\author[1]{Schlauch, Tobias \orcidlink{0000-0001-8760-8913}}

\affil[1]{Deutsches Zentrum für Luft- und Raumfahrt e.\,V.}
\affil[2]{Forschungszentrum Jülich GmbH}
\affil[3]{Helmholtz-Zentrum Dresden-Rossendorf e.\,V.}
\affil[*]{Corresponding Author (\href{mailto:team@software-metadata.pub}{team@software-metadata.pub})}

\makeglossaries
\longnewglossaryentry{ci-cd}
{
    name=CI/CD
}{
    Describes continuous integration/continuous deployment solutions, usually integrated in version control platforms
    to run automated workflows triggered by changes uploaded to the version control service. These workflows are
    primarily used to run automated software tests, but can also be used to run any other software automatically.
    Examples for CI/CD tools include GitLab CI, GitHub Actions and Jenkins automation servers.
}
\longnewglossaryentry{sw-package}{
    name=Software Package,
    text=software package
}{
    Describes a unit of functionally and/or semantically self-contained software. This meaning is opposed to the notion
    of package in some programming language, e.g., Java, where it is used to signify the namespace of a smaller unit,
    e.g., a source code file or a class. Other terms for software package include: (software) product, software (sg.),
    piece of software.
}
\longnewglossaryentry{dev-platform}{
    name=Software Development Platform
    text=software development platform
}{
    means an online platform that supports the software development process through the combination of a version-controlled source code repository and additional tools such as issue trackers, code review tools, automation pipelines, etc. Popular examples are GitHub and GitLab.
}
\newglossaryentry{sourcerepo}{
    name=Source Code Repository,
    text=source code repository,
    plural=source code repositories,
    description={is a version controlled storage of directories and files usually as part of a software development platform}
}
\newglossaryentry{pubrepo}{
    name=Publication Repository,
    text=publication repository,
    plural=publication repositories,
    description={A public catalogue of published artifacts that contains both the artifacts themselves as well as standardized metadata for the artifact. Each artifact is addressable with a unique identifier}
}
\newglossaryentry{repostatus}{
    name=Repository Status Controlled Vocabulary,
    text=repository status,
    plural=repository states,
    description={Based on the terminology from \url{https://www.repostatus.org}, we use the different stati throughout this paper. Stati involve \repostatus[label]{Concept}, \repostatus[label]{WIP}, \repostatus[label]{Suspended}, \repostatus[label]{Abandoned}, \repostatus[label]{Active}, \repostatus[label]{Inactive}, \repostatus[label]{Unsupported} and \repostatus[label]{Moved}}
}
\newglossaryentry{webhook}{
    name=Webhook,
    text=webhook,
    plural=webhooks,
    description={Common web technique: some software sending an HTTP POST request to a target system with the intent to trigger some kind of reaction. The request may carry a (JSON) payload, containing context, authentication, parameters and other information}
}
\newglossaryentry{static-metadata}{
    name=Statically available metadata,
    text=statically available metadata,
    plural=statically available metadata,
    description={Statically available software metadata can be accessed from static sources such as dedicated files or parts of files, version control systems or other forms of repositories, \gls{ci-cd} contexts, file systems or platform APIs}
}

\begin{document}
\maketitle
\thispagestyle{empty}

\begin{abstract}\label{abstract}
\addsec{Abstract}
To satisfy the principles of FAIR software, software sustainability and software citation, research software must be formally published.
Publication repositories make this possible and provide published software versions with unique and persistent identifiers.
However, software publication is still a tedious, mostly manual process.

To streamline software publication, HERMES, a project funded by the Helmholtz Metadata Collaboration, develops automated
workflows to publish research software with rich metadata.

The tooling developed by the project utilizes continuous integration solutions to retrieve, collate, and process existing metadata
in source repositories, and publish them on publication repositories, including checks against existing metadata requirements.
To accompany the tooling and enable researchers to easily reuse it, the project also provides comprehensive documentation and
templates for widely used CI solutions. In this paper, we outline the concept for these workflows, and describe how our solution
advance the state of the art in research software publication.
\end{abstract}

\vspace*{\fill}
\begin{center}
    \footnotesize
    This work is licensed under \href{https://creativecommons.org/licenses/by/4.0/}{\faCreativeCommons\faCreativeCommonsBy\,4.0}.    
\end{center}
\clearpage

\begin{tcolorbox}[colback=blue!5!white,colframe=blue!75!black,title=Requesting community feedback]
\paragraph{Community feedback}\label{community-feedback}

At this stage, we would like to reach out to the community to gain insights into what desirable solutions can 
look like, what potential pitfalls are, etc.

We kindly ask readers and other interested parties to provide feedback on the concept detailed here via its PubPeer page at
\url{https://software-metadata.pub/concept-paper-community-reviews}.
\tcblower

\begin{itemize}  
    \item What metadata types, formats and standards are missing from our lists in \nameref{subsec:metadata}?
    \item Does table \ref{tab:comparision-metadata-types}, mapping metadata types to metadata formats, miss anything important, or misrepresent something?
    \item Does table \ref{tab:workflows} miss any other publication workflows you know and should be included?
    \item What are desired additional outputs of the automated workflows in addition to deposition in the targeted
          publication repositories? E.g., would the creation of \nameref{par:metadata-formats-codemeta} or 
          \nameref{par:metadata-formats-cff} be helpful, are there other desired output types?
    \item Should the HERMES pipelines leverage workflow domain specific languages (such as
          \href{https://www.commonwl.org/}{Common Workflow Language (CWL)}, \href{https://nextflow.io/}{Nextflow}, etc.)
          to knit together existing tools, such as harvesters (e.g., \nameref{par:tooling-somef}), 
          converters (e.g., \nameref{par:tooling-cffconvert}), and deposition tools (e.g., \nameref{par:workflows-zenodraft})?
    \item While we initially restrict the scope of HERMES with regard to metadata validation to linting 
          (see \ref{subsec:concept-scope}), there may be other factors that influence the validity of metadata.
          It is known that VCS contributors metadata, for example, is unsuitable to be used as valid authorship
          metadata, as there may be people qualifying as authors who have not contributed to a source code repository,
          or people that are contributors but do not qualify as authors under a given definition of authorship.
          Therefore, other metadata sources, such as files in the Citation File Format may be more trustworthy.
          What are other pitfalls concerning the validity of metadata that HERMES should be aware of?
\end{itemize}

\end{tcolorbox}
\clearpage

\section{Introduction}\label{sec:introduction}
There is increasing awareness that software is a valid research output and should be treated as such~\cite{jaySoftwareMustBe2021,SoftCitePrinciples}. Thus, software is increasingly published in public repositories or software journals~\cite{TrackingCitations}. This is a necessary step in transferring the FAIR principles to software~\cite{FAIR4RS-Principles} -- i.e., for finding, understanding, reusing, sharing and citing software -- and promoting it to first class research citizenship. Recent policy updates from universities, research institutes (e.g., at Helmholtz Association (HGF)~\cite{HelmholtzSWPolicy}) and funders such as the Deutsche Forschungsgemeinschaft (DFG)~\cite{DFG-GRP19} reflect this progress. Metrics for published software may inform funding decisions in the future.

The main driver for a fulfillment of the functions of FAIR software is software metadata~\cite{FAIR4RS-FreshLook}, and thus, publication of research software with rich metadata is essential. In modern research software development, metadata on different software properties is created automatically, semi-automatically or manually at different stages, and in different places and formats. While this metadata can already be collected, verified and validated, and edited to be published with software, there is currently no streamlined, automated process or workflow to do so. This in turn disincentivizes the researchers, research software engineers and maintainers who create and maintain research software, to publish it with rich metadata.

In this paper, we describe a concept for automated publication of research software with rich metadata via existing automation tools. The concept is being developed in the project HERMES (\textbf{HE}lmholtz \textbf{R}ich \textbf{ME}tadata \textbf{S}oftware Publication), conducted at the German Aerospace Center (DLR), Forschungszentrum Jülich (FZJ) and Helmholtz Zentrum Dresden Rossendorf (HZDR), and funded by the Helmholtz Association of German Research Centers‘ Helmholtz Metadata Collaboration (HMC) initiative (see \nameref{acknowledgments}).

The work packages related to the concept as described in this document yield a number of outputs: 

\begin{itemize}
  \item \textbf{software} to retrieve existing software metadata in source code repositories, collate and process them to produce a coherent set of metadata for the current state of a given repository;
  \item \textbf{templates}, e.g. for \gls{ci-cd} systems and workflow engines, to run the metadata tools on a users' source repositories;
  \item \textbf{documentation} and examples for the outputs the project provides.
\end{itemize}

In the following sections, we describe the state of the art for software metadata, available tooling to work with metadata and existing approaches to automation. We then lay out our concept for research software deposits with rich metadata in automated workflows for two target platforms. In the process, we specify requirements for the structure and contents of source code repositories, define an iterative process for the inclusion of increasingly unstructured metadata, detail the scope of the project and provide a high-level outline of the implementation of both interfaces and tooling. 

We request community feedback for the concept detailed here (see \crtlnameref{community-feedback} above). Based on feedback we receive, the HERMES project partners design interfaces and develop software tools to automatically aggregate metadata included in software repositories. The interfaces and software tools combination provide an extensible, \gls{ci-cd}-driven serverless solution that enables direct ingestion into publication repositories, such as Zenodo or Harvard Dataverse, and other repositories using the underlying \nameref{par:repos-inveniordm} or \nameref{par:repos-dataverse} software.

\clearpage
\section{State of the art}\label{sec:state-of-the-art}

\subsection{Metadata}\label{subsec:metadata}
Software metadata provides information about software, or specific properties of software. It is created intentionally, or generated as a side-product during software development processes. As such, metadata can also pertain to different parts of software, or a \gls{sw-package} as a whole. Additionally, it can exist in different modes, i.e., as structured or unstructured metadata. It may also pertain to different aspects of the software, e.g., the license regulating its use and development, its creation context, etc. Consequently, there exist different types of metadata.

Software metadata may be provided in dedicated files, or as part of some file. Metadata can also be part of version control systems
or other forms of repositories, as well as the file system or platform hosting the source code. We define these as \gls{static-metadata}.

Structured metadata, especially when persisted in files, may furthermore come in specific formats. These may be standard formats through formal processes or community practice.

In this section, we describe different types of software metadata, as well as formats they are provided in. Table \ref{tab:comparision-metadata-types} on page \pageref{tab:comparision-metadata-types} shows a mapping of metadata types to the formats they are commonly provided in, based on our experience.

\subsubsection{Types}\label{subsubsec:metadata-types}
There are generic types of metadata that exist for software but may also be found for other digital data, as well as software-specific metadata.
The latter is partly due to the fact, as described in \cite{PeerJSoftVsDataCitation}, that software is both static (as source code, i.e., digital data)
and dynamic (at runtime).

\begin{multicols}{2}
    \raggedright
    Generic software metadata includes:
    \begin{itemize}  
        \item Software name
        \item File system metadata \\ {\small (e.g., file sizes, number of files, etc.)}
        \item Authorship and contributorship information
        \item Reference to the documentation pertaining to the software
        \item Legal and licensing information
        \item Funding information
        \item Domain context
        \item Citation metrics
        \item Location metadata \\ {\small (e.g., download or instance URLs, etc.)}
        \item Publication dates, etc.
        \item Categorization information \\ {\small (e.g., application category, keywords, etc.)}
        \item Availability information \\ {\small (e.g., purchasing costs, etc.)}
        \item Identifiers
        \item Relational metadata \\ {\small (e.g., software \textit{is part of} another work)}
        \item High-level description
    \end{itemize}
    \columnbreak
    
    Software-specific metadata includes:
    \begin{itemize}  
        \item Dependency information
        \item Lines of code
        \item Programming language
        \item Version information \\ {\small (e.g., metadata from version control systems, publication platforms, or even file names; version identifiers, etc.)}
        \item Runtime requirements, including hardware requirements
        \item References to work the software is built on, or relates to
        \item Software metrics \\ {\small (e.g., quality metrics like code coverage, ...)}
        \item Development metrics \\ {\small (e.g., pertaining to issues, pull requests, ...)}
        \item Usage metrics \\ {\small (e.g., downloads, stars, citations, ...)}
        \item Infrastructural metadata \\ {\small (e.g., build and CI systems used to produce version artifacts, etc.)}
    \end{itemize}
\end{multicols}

\subsubsection{Formats}\label{subsubsec:metadata-formats}
Some metadata are persisted in files that have specific formats, or are integrated in specific formats as part of other files.
Such files are usually persisted and version controlled alongside source code. 

Other metadata must be retrieved from third-party systems, e.g., the version control system, source code platform
(GitHub, GitLab, or other), etc., if available. Some are only available on other platforms or systems and may not be
retrievable from the source code repository.

\paragraph{Plain text files}\label{par:metadata-formats-plaintext}
Some software metadata is provided in plain text. There are some typical dedicated plain text files, such as license files
(\fn{LICENSE}, \href{https://reuse.software/spec/}{REUSE Specification} compliant files, etc.) or
citation metadata files (plain text \fn{CITATION} files), that can reasonably be assumed to contain only relevant metadata. 

Other files mix metadata and other information, such as documentation files (\fn{README} files and other plain text documentation),
community files containing contributor information, a code of conduct, governance information, or other information.
Other relevant metadata may be provided in plain text in a less overt manner, e.g., embedded in source code files. 

Generally, while plain text metadata is easily accessible for human readers, they are perhaps the hardest to process using
automated approaches. This is due to their less structured form and a lack of clarity with regard to semantics. A plain 
\fn{CITATION} file, for example, may provide retrievable publication metadata, but there is no way of automatically verifying
that the metadata unambiguously pertains to a specific version of the software it is provided with, or indeed something
else entirely. 

The same is true for any plain text metadata and specifically for metadata that are embedded in plain text: while methods
exist to extract metadata from plain text, they rely on heuristics that can only produce assumptions with some degree of
confidence in their significance and correct categorization.

\paragraph{schema.org files}\label{par:metadata-formats-schema-org}
schema.org~\cite{SchemaOrg} provides schemas for structured data markup. These are commonly used in HTML to provide metadata that is reused by search engines. schema.org schemas are also used as basis for metadata in RO-Crate~\cite{RO-Crate}. Additionally, there is ongoing work\footnote{\url{https://github.com/codemeta/codemeta/issues/232}} to add missing terms from the CodeMeta schema~\cite{CodeMetaSchema} to schema.org.

\paragraph{CodeMeta files}\label{par:metadata-formats-codemeta}
CodeMeta~\cite{CodeMetaSchema} is a format for generic software metadata, implemented in JSON-LD, extending \nameref{par:metadata-formats-schema-org}. It is used to provide comprehensive information about software, with some focus on academic use cases.

\paragraph{Citation File Format}\label{par:metadata-formats-cff}
The Citation File Format~\cite{CffSchema} is a format to provide citation metadata for software, implemented in YAML. Its focus is exclusively to provide citation-relevant metadata in a form that is both human- and machine-readable.

\paragraph{Zenodo JSON files}\label{par:metadata-formats-zenodo-json}
The open access publication platform Zenodo~\cite{Zenodo} uses its own metadata schema, implemented in JSON\footnote{\url{https://developers.zenodo.org/\#github}}. The schema is used in practice to provide metadata for works that are being published on Zenodo, e.g., through the GitHub-Zenodo integration (see also \nameref{par:workflows-pull} and table \ref{tab:workflows} on page \pageref{tab:workflows}).

\paragraph{\hologo{BibTeX} files}\label{par:metadata-formats-bibtex}
\hologo{BibTeX} files contain citation metadata for one or more works. The format is standardly used as input for citations and bibliographies
in \LaTeX\ documents, but is sometimes also adapted to provide convenient citation metadata for software, for example in a dedicated file in
\glspl{sourcerepo} (sometimes called \fn{CITATION}), or embedded in a text or marked up document such as a \fn{README} file. 

In the context of metadata retrieval, \hologo{BibTeX} files or snippets pose the same challenges as plain text due to their generic nature.
It is hard to understand if a \hologo{BibTeX} item is describing the software package it is provided with, one of its versions, or something
else entirely. The biblatex-software~\cite{sw:biblatex-software} package for \LaTeX\ solves issues around citing different software reference
types in scholarly publications, but the general issue with \hologo{BibTeX} items and their relation to given software remains.

\paragraph{Manifest files}\label{par:metadata-formats-manifests}
Manifest files describe a \gls{sw-package} or a subunit of a software package. They exist for different programming languages and frameworks and come in a variety of implementation formats. Some examples include Project Object Model (POM) XML files for Java projects using the Apache Maven build management tool, JAR manifests for packaged Java applications, or \fn{setup.py}/\fn{pyproject.toml} files that contain metadata for Python packages built with Python’s distutils.

\paragraph{Configuration files}\label{par:metadata-formats-config}
Configuration files are often added to source code repositories to leverage third-party tools such as \gls{ci-cd} services or source code or documentation generators. They come in different formats, of which common ones are plaintext key-value definitions, INI, TOML, YAML, JSON and XML. Configuration files may include metadata pertaining to, e.g., software dependencies, development and publication processes, etc. One popular example for configuration files that contain contribution metadata are those for the All Contributors\footnote{\url{https://allcontributors.org}} specification.

\paragraph{Linked Data files}\label{par:metadata-formats-linked-data}
Linked data provides an extensible way to describe research software in depth. Common linked data formats are based on RDF~\cite{W3C-RDF}, using serializations like JSON-LD~\cite{W3C-JSON-LD} or Turtle~\cite{W3C-RDF-Turtle}. They express not just attributes or simple relations, but reuse formalized concepts (ontologies) to describe usage, ideas, context, etc., in much greater depth than other formats listed above.

Although ontologies are a powerful concept, there does not seem to be sufficient uptake of using software ontologies to describe software in practice. This may change in the future.

Commonly known ontologies for software include Software Description Ontology~\cite{SoftDescOnto}, Software Ontology~\cite{SoftwareOntology}, SEON~\cite{SeonOntology} and DOAP~\cite{DoapOntology}.

\paragraph{Version control system}\label{par:metadata-formats-vcs}
The version control system in use may provide valuable metadata from the commit history. Especially in distributed version control systems like Git, 
the complete history is available locally. These metadata are part of a project context and not formalized in a file, but may be retrieved by using command
line interfaces or wrapping libraries of the version control system at hand. They may provide insights for example into contributors, file metadata,
changes over time or related software/data.

\paragraph{Platform API responses}\label{par:metadata-formats-apis}
While not strictly a format, software metadata is also available from different (e.g., REST, GraphQL) APIs, such as those for querying software
development platforms (like GitHub or GitLab), WikiData, or publication repositories. These may provide metadata on software development
processes, version control, contributions, publications, as well as other metadata. These APIs can usually be queried through their
endpoints -- or globally using a common query language such as~\cite{W3C-SPARQL} -- and responses are usually provided as JSON or XML
that can easily be persisted and reused.

\subsection{Standards}\label{subsec:metadata-standards}
As of now, there are no software metadata formats relating to our work that are formally standardized and cover HERMES’
scope completely. Some de facto standards exist:

\begin{itemize}
    \item CodeMeta’s ongoing integration\footnote{\url{https://github.com/codemeta/codemeta/issues/232}} into schema.org
          promises at least future de facto standardization. 
    \item The DataCite Metadata Schema~\cite{DataCiteSchema}, although used widely, has a much more generic scope again
          than, e.g., CodeMeta, and does not implement as large a vocabulary pertaining to software 
          (see the CodeMeta-DataCite crosswalk\footnote{\url{https://codemeta.github.io/crosswalk/datacite/}}).
\end{itemize}

Standards in similar stages exist in related areas, such as for research objects in general: 
\begin{itemize}
    \item RO-Crate~\cite{RO-Crate} packages research objects with their metadata, its schema based on schema.org.
    \item BagIt~\cite{BagItRFC} is a generic packaging standard used in, e.g., libraries, to package file structures with
          metadata, where the metadata schema is very generic and not targeted to software specifically.
\end{itemize}

\subsection{Integrations}\label{subsec:metadata-integrations}
Some platforms and tools provide integrations for some of the above-mentioned formats. As opposed to concrete (software)
tools for working with formats that are available to end-users, integrations are embedded in their hosts and are not
directly addressable by users.

\paragraph{Plain text}\label{par:metadata-integrations-plaintext}
Many platforms support metadata provided in plain text or light\-weight markup languages by rendering them for presentation
to end users. This often includes detecting URLs and converting them to HTML hyperlinks. Examples include Markdown
rendering on, e.g., GitHub, GitLab, and many other platforms.

\paragraph{schema.org}\label{par:metadata-integrations-schemaorg}  
RO-Crate~\cite{RO-Crate} uses schema.org~\cite{SchemaOrg} schemas to record and provide core metadata. Both \nameref{par:repos-dataverse} and
\nameref{par:repos-inveniordm} offer metadata exports as schema.org JSON-LD.

\paragraph{CodeMeta}\label{par:metadata-integrations-codemeta}
Software Heritage~\cite{SWHArchive} uses a subset\footnote{\url{https://docs.softwareheritage.org/devel/swh-indexer/metadata-workflow.html\#supported-codemeta-terms}}
of the CodeMeta vocabulary to map intrinsic metadata formats discovered in source repositories.
CaltechDATA ingests CodeMeta files to create metadata records from the included metadata~\cite{ASCL-CaltechAMES}~\cite{sw:AMES}.
The Astrophysics Source Code Library (ASCL) produces pre-filled Code\-Meta files from its records~\cite{ASCL-Software-Citation} --
this functionality is accessible to users through appending the URL for a record page with \fn{/codemeta.json}.

\paragraph{Citation File Format}\label{par:metadata-integrations-cff}
\begin{itemize}  
    \item Via its Github-Zenodo integration, Zenodo ingests the metadata from \fn{CITATION.cff} files and uses them to populate their
          metadata records~\cite{sw:invenio-github} (see also table \ref{tab:workflows} on page \pageref{tab:workflows}).
    \item Via its connector browser plugins, Zotero ingests \fn{CITATION.cff} files discovered in source code repositories and saves
          the metadata in its internal format~\cite{FennerBlogCff}.
    \item JabRef can import \fn{CITATION.cff} files (feature merged, currently awaiting release\footnote{\url{https://github.com/JabRef/jabref/issues/7945}}).
    \item The Astrophysics Source Code Library (ASCL) produces pre-filled \fn{CITATION.cff} files from its records. This functionality
          is accessible to users through appending the URL for a record page with \fn{/CITATION.cff}~\cite{ASCL-Software-Citation}.
    \item GitHub provides a template\footnote{\url{https://docs.github.com/en/repositories/managing-your-repositorys-settings-and-features/customizing-your-repository/about-citation-files}}
          for creating new \fn{CITATION.cff} files via its UI and ingests existing \fn{CITATION.cff} files, extracts the metadata, converts
          them to a citation style and BibTeX, and provides them in a widget for end users to copy~\cite{FennerBlogCff}. 
          There is a feature request to implement the same for GitLab\footnote{\url{https://gitlab.com/gitlab-org/gitlab/-/issues/337368}}.
    \item An extension\footnote{\url{https://www.higithub.com/citation-file-format/issue/citation-file-format/356}} to the Sphinx platform 
          for documentation rendering ingests an existing \fn{CITATION.cff} file and, converts it to different citation formats, and provides 
          it to end users in a widget for copying (see example in~\cite{CffSphinxExample}).
    \item Software Heritage can ingest \fn{CITATION.cff} as intrinsic metadata format and map it
          (see \href{https://docs.softwareheritage.org/devel/swh-indexer/metadata-workflow.html}{SWH Indexer Metadata Workflow},
          referenced in~\cite{FennerBlogCff}).
\end{itemize}

\paragraph{Zenodo JSON}\label{par:metadata-integrations-zenodo-json}
Via its GitHub-Zenodo integration, Zenodo ingests\footnote{\url{https://developers.zenodo.org/\#add-metadata-to-your-github-repository-release}}
\fn{.zenodo.json} files and populates metadata records from the provided metadata~\cite{sw:invenio-github}.

\paragraph{BibTeX}\label{par:metadata-integrations-bibtex} 
Usually, the metadata from \hologo{BibTeX} \fn{.bib} files are converted into string formatted in a citation style and displayed 
as citations and items in the references list of \LaTeX\ based documents.

biblatex-software~\cite{sw:biblatex-software} is a reference biblatex implementation of a bibliography style extension
that includes software-specific \hologo{BibTeX} entries and integrates these metadata in \LaTeX\ documents.

\paragraph{Manifest files}\label{par:metadata-integrations-manifests}
Some or all information from manifest files is rendered on package/artifact repositories’ sites for the respective package,
e.g., on \href{https://mvnrepository.com}{Maven Central}, \href{https://pypi.org}{PyPI}, 
\href{https://www.npmjs.com}{NPMJS}, \href{https://www.debian.org/distrib/packages}{Debian Packages} etc.

\subsection{Tooling}\label{sec:tooling}
The following section is an attempt to gather tools available for metadata extraction, collection and durable publication.
Please note: section \ref{subsubsec:workflows-blocks} contains any tooling to be used as building blocks
for (automatable) workflows around software depositions. These lists may not necessarily be complete.

\subsubsection{Metadata}\label{subsec:tooling-metadata}
Existing toolsets for metadata may operate on different stages of metadata presence. The spectrum starts at zero 
prior information and requires extraction from arbitrary text files and context information. Moreover, it can be 
accomplished by reusing software package metadata via so called crosswalks, or even  conversions between different
metadata formats. Some may give users a hand to create well-structured metadata from the start.

\paragraph{Software Metadata Extraction Framework (SoMEF)}\label{par:tooling-somef}
SoMEF~\cite{sw:SoMEF}~\cite{SoMEF} extracts data from \fn{README} text files and
other files that may include metadata using a neural network. It may also retrieve details from \gls{dev-platform}
such as GitHub, using their APIs. It creates, e.g., CodeMeta JSON-LD or Turtle RDF files using 
The Software Ontology~\cite{SoftDescOnto}. The projects \gls{repostatus} is \repostatus[label]{active}.

\paragraph{CaltechDATA Automated Metadata Service (AMES)}\label{par:tooling-ames}
AMES~\cite{sw:AMES} may be used to create and update scholarly output records in services like
CaltechDATA, highly specific for Caltech and based on \ref{par:repos-inveniordm}. For software publications
a Python script to update \nameref{par:repos-inveniordm} records from \nameref{par:metadata-formats-codemeta}
files is in use~\cite{ASCL-CaltechAMES}. The projects \gls{repostatus} is \repostatus[label]{active}.

\paragraph{codemeta2cff GitHub Action}\label{par:tooling-codemeta2cff}
The codemeta2cff GitHub Action~\cite{sw:codemeta2cff} provides automatic conversion from a
\nameref{par:metadata-formats-codemeta} file to a \nameref{par:metadata-formats-cff} file in a GitHub Action.
The projects \gls{repostatus} is \repostatus[label]{suspended}.

\paragraph{CodeMeta Crosswalks}\label{par:tooling-crosswalks}
CodeMeta Crosswalks~\cite{sw:codemeta-crosswalks} are a set of comma-separated value (CSV) files, containing
two columns. Each row depicts a metadata field in \nameref{par:metadata-formats-codemeta} and a
corresponding field in some other schema. 
While not an executable tool, these crosswalks describe a mapping with limited interoperability, as most
other standards aren’t as detailed.
The projects \gls{repostatus} is \repostatus[label]{inactive}.

\paragraph{CodeMeta Generator}\label{par:tooling-codemeta-gen}
CodeMeta Generator~\cite{sw:codemeta-generator} is a Javascript-based web UI to help you create 
\nameref{par:metadata-formats-codemeta} for inclusion in your software repository. The projects
\gls{repostatus} is \repostatus[label]{inactive}.

\paragraph{Citation File Format Converter}\label{par:tooling-cffconvert}
cffconvert~\cite{sw:cffconvert} is a Python based command line tool to transform a given 
\nameref{par:metadata-formats-cff} file into other destination formats like \nameref{par:metadata-formats-bibtex},
\nameref{par:metadata-formats-codemeta} and others. It is also available as a GitHub Action.
The projects \gls{repostatus} is \repostatus[label]{active}.

\paragraph{Citation File Format Initializer}\label{par:tooling-cffinit}
cffinit~\cite{sw:cffinit} is a Javascript based interactive web form to assist you in creating a 
new or updating an existing \nameref{par:metadata-formats-cff} file in your browser.
The projects \gls{repostatus} is \repostatus[label]{active}.

\subsubsection{Publication repositories}\label{subsubsec:repositories}
\Glspl{pubrepo} are public catalogue containing publications of digital artifacts together with the metadata describing them.
Usually, publication repositories provide landing pages for each artifact, including versions of the same object.
As such, they are different from registries, that usually focus on the collection of metadata and their presentation.
They are also different from archives, that focus on long-term archival of artifacts only. Additionally there may exist
differences in how records are added to publication repositories, registries and archives.

One of the main advantages of publication repositories is that they enable a combination of discovery of digital objects
through their metadata, and direct access to the object artifacts themselves. HERMES focuses on publication repositories
exclusively, and specifically on two publication repository software projects as deposition targets, for the reason that
they represent commonly used platforms both within the Helmholtz Association and beyond: Dataverse project and InvenioRDM.

Research software is also represented in digital preservation archives (like the Software Heritage Archive~\cite{SWHArchive}),
catalogues and directories (like the Research Software Directory~\cite{RSD-2018}). Targeting these platforms may be a 
future direction of development for HERMES, see also \ref{subsubsec:concept-scope-future}.

\paragraph{Dataverse project}\label{par:repos-dataverse}
The \href{https://dataverse.org}{Dataverse project} is an open source repository software.

Its focus is currently on providing services for \href{http://purl.org/dc/dcmitype/Dataset}{Dataset} deposition, although 
software may be deposited as part of datasets, too. No built-in support for software metadata schemas, software licenses
or propagating software metadata to PID registrars is available.

Despite versioning support for datasets, neither integrating software release versioning is available nor support for
software citations as a concept and individual releases.

The Dataverse community runs a working group for software, workflow and container related topics.
Its website can be found at \url{https://swc.wgs.gdcc.io} The projects \gls{repostatus} is \repostatus[label]{active}.

\paragraph{InvenioRDM}\label{par:repos-inveniordm}
The \href{https://inveniosoftware.org/products/rdm/}{InvenioRDM project} has the goal to provide a turn-key research data
management repository based on Invenio Framework and Zenodo. The publication of special software datasets is possible,
but the standard set of metadata for the description of Invenio records is used as well as in \ref{par:metadata-formats-zenodo-json}.

Invenio provides versioning, DOI registration and supports multiple data types for the publication such as Publication, Poster,
Presentation, Dataset, Image, Video/Audio, Software, Lesson, Physical objects or Other 
(list\footnote{See \texttt{upload\_type} at \url{https://zenodo.org/schemas/deposits/records/legacyrecord.json}} taken from Zenodo).
No builtin support for software metadata schemas is available at the moment. 

For a software publication the use of optional webhooks can be used as introduced in \nameref{par:workflows-pull}.

The projects \gls{repostatus} is \repostatus[label]{active}.

\subsection{Workflows}\label{subsec:workflows}
Depositing scientific software to archives like Software Heritage~\cite{SWHArchive} or \crtlnameref{subsubsec:repositories}
requires some kind of workflow, involving manual steps or automation.

Workflows may be coarsely categorized into \enquote{push} or \enquote{pull} based approaches. Both have their pros 
and cons, but within the context of scholarly software publications, push-based approaches have the advantage of 
not having to expose the \gls{sourcerepo} to the publishing service. As making a software FAIR does not require 
code access~\cite{FAIR4RS-Principles}\cite{FAIR4RS-FreshLook}, this may be beneficial to increase the number of 
software publications even for closed source research software.

Table \ref{tab:workflows} on page \pageref{tab:workflows} provides an in-depth overview of known full-fledged workflows
and tools used within custom automated jobs pertaining to software publications. Both categories of workflows are
covered and analysed for their metadata extraction and publication capabilities. Please note: neither the table nor
the following list of building blocks need to be complete.

\paragraph{Pull-based workflows}\label{par:workflows-pull} may be subdivided into \enquote{harvesting} and \enquote{triggered} types.

\enquote{Harvesting} for new or changed datasets is an often used pattern within the world of text and data publications.
The well-known OAI-PMH is used for inter-repository talk, while harvesting in the context of workflows is attached
to pulling commits from public accessible \glspl{sourcerepo}. Retrievals may be scheduled or kicked off by some event.
To provide an example: both the Software Heritage Archive~\cite{SWHArchive} and the Research Software Directory~\cite{RSD-2018}
use scheduled harvesting: they check for changes in source code repositories, publishing repositories, etc. and incorporate them, which 
might involve updating existing metadata.

Using a \gls{webhook} to \enquote{trigger} some action is a well known technique in distributed systems. 
Within the publication business a webhook may even trigger a harvesting action with certain parameters like a target.
The GitHub-Zenodo integration\footnote{\url{https://guides.github.com/activities/citable-code}} is a good example for this,
sending a webhook request on \gls{sw-package} releases (tagged revisions within a \gls{sourcerepo}) to trigger the harvest.

\paragraph{Push-based workflows}\label{par:workflows-push}
While pull-based workflows have their advantages, in some scenarios you might want to publish your software in a more active fashion.

Push-based workflows gather metadata and/or artifacts and deposit them via some API endpoint into a service like a repository,
registry or archive. They may also split certain tasks across different services, which may be harder to achieve with
common pull-based workflows.

Examples for push-based workflows are even harder to find than pull-based workflow approaches. This might be due to the
convenience of commonly known pull-based workflows and the not-yet popular task of software publications. Please let us
know of any prototypical example we haven’t listed in table \ref{tab:workflows} on page \pageref{tab:workflows} yet.

\subsubsection*{Building Blocks}\label{subsubsec:workflows-blocks}
The following tools provide building blocks to create an automatable depositing workflow, interfacing with target repositories, registries or archives.

\paragraph{Zenodraft}\label{par:workflows-zenodraft}
Zenodraft~\cite{sw:zenodraft} and the corresponding GitHub action~\cite{sw:zenodraft-action} may be used to draft, push and publish
new deposits on \href{https://zenodo.org}{Zenodo}, a commonly known general purpose repository based on \nameref{par:repos-inveniordm}.
New and existing deposits may be enhanced with metadata via \nameref{par:metadata-formats-zenodo-json}.
The project's \gls{repostatus} is \repostatus[label]{suspended}.

\paragraph{Software Heritage Github Action}\label{par:workflows-swh-gh-action}
SWH Github Action~\cite{sw:swh-save-action} acts like a webhook: it sends a archive request to an Software Heritage Archive 
API endpoint with the repository to archive given as parameter.
Software Heritage Archive services take care\footnote{See \url{https://docs.softwareheritage.org/devel/swh-indexer/metadata-workflow.html}}
of reading a \nameref{par:metadata-formats-codemeta}, \nameref{par:metadata-formats-cff} or other metadata files via 
\nameref{par:tooling-crosswalks} (if included) to add metadata to its archive.
The project's \gls{repostatus} is \repostatus[label]{suspended}.

\paragraph{Software Heritage Deposit Command Line Tool}\label{par:workflows-swh-deposit-cli}
The CLI part of a SWORD~\cite{SWORD} based deposition service~\cite{sw:swh-deposit-cli} 
for the Software Heritage Archive may be used to actively push metadata and/or artifact ingests.
It seems not to be intended for public usage but integrations with software/data repositories.
The project's \gls{repostatus} is \repostatus[label]{active}.

\paragraph{Dataverse Uploader GitHub Action}\label{par:workflows-dataverse-action}
The Dataverse Uploader Github Action~\cite{sw:dataverse-uploader-action} enables uploading content from a \gls{sourcerepo} into
a \enquote{dataset} on a target \nameref{par:repos-dataverse} installation. It allows to replace all or add to files and their metadata.
The action may also publish a new dataset version afterwards. The projects \gls{repostatus} is \repostatus[label]{active}.

\clearpage
\section{Concept}\label{sec:concept}

\subsection{Overview}\label{subsec:concept-overview}
The coupling of software development platforms with publication repositories for automated data exchange is a first and important step
for easier and automated software publications. This technical foundation also immediately raises the question as to the requirements
towards a source code repository on a development platform to benefit from such automation. Current first generation tools simply copy
the content of a source code repository to create or update a publication repository entry. The metadata of the publication repository
entry can originate from a special file in the source code repository.

Questions addressed by HERMES that extend the status quo are:

\begin{itemize}  
    \item How to automate collation of metadata from different sources for automated publication?
    \item How to treat different components in a source code repository (software, documentation and data)?
    \item How to deal with source code repositories that contain more than one software product?
    \item How to deal with publication of executable software artifacts generated from the software repositories?
    \item How to enable closed source but FAIR software~\cite{FAIR4RS-FreshLook} for these processes?
    \item How to synchronize existing metadata automatically after publication?
\end{itemize}

\subsection{Source code repositories}\label{subsec:concept-sourcerepos}
HERMES targets both GitHub and GitLab as software development platforms as they are the de facto community standard and are widely used
both as public cloud services and on-premise installations. Both services offer an API for interaction with other services and, thus,
provide an ideal starting point for HERMES to add on to the existing solutions.

\subsubsection{Different ways of setting up software projects}\label{subsubsec:concept-sourcerepos-ways}
A source code repository containing only one software package is an easy, common case and straightforward in publication. Integrated
software documentation as part of the repository is considered a part of the software package and does not need to be treated differently.

For other cases, e.g., data alongside the source code or multiple software packages in one source code repository, HERMES allows the user to specify which parts of the repository to include in a publication.

\subsection{Metadata sources}\label{subsec:concept-metadata-sources}
HERMES decidedly does not limit the metadata sources it works on to specific types. However, implementation follows an iterative
process, with support for different types being added in stages.

\begin{enumerate}  
    \item Firstly, we collect structured metadata that can be tested for availability, e.g., dedicated metadata files, such as
          \fn{codemeta.json}, \fn{CITATION.cff}, \fn{LICENSE} files , version control metadata ,
          software development platform metadata (all described in \nameref{subsubsec:metadata-formats}).
    \item Secondly, we attempt to mine structured data that may or may not contain relevant metadata, e.g., manifest files, configuration files, etc.
    \item Finally, we attempt to mine unstructured metadata from, e.g., plain text files.
\end{enumerate}

In cases where several instances of metadata sources exist and contain overlapping -- and potentially diverging --
information, we follow a set of heuristics to defensively establish source precedence and avoid conflicts or bad metadata.

\subsection{Scope}\label{subsec:concept-scope}
HERMES aims to enable software publications in publication repositories where metadata is transported from the source code repository
to the publication repository. During this first iteration of the project, it interfaces with two popular publication repository 
software products: the \nameref{par:repos-dataverse} and \nameref{par:repos-inveniordm}.

\subsubsection{Out of scope}\label{subsubsec:concept-scope-out}
The tooling developed and provided needs to be shaped in scope and size. Thus, at this stage HERMES

\begin{itemize}  
    \item does not offer license compatibility checks,
    \item does not resolve values from external vocabularies (e.g., WikiData, triplestores using software metadata ontologies) or
          other persistently identified resources (ORCID, ROR, other publications, ...),
    \item does not validate metadata beyond pure linting functionality,
    \item does not create provenance, workflow or pipeline models (but may be a part of these),
    \item does not search or resolve publications of (software) dependencies and
    \item does not run software being published to collect runtime-specific metadata.
\end{itemize}

The software and reusable workflow templates that HERMES provides does not constitute, or be provided as, a \enquote{service}
(web service, REST API, SaaS, PaaS, IaaS, etc.) or other infrastructure component.

Instead, research software projects can reuse the solutions in their own software projects, e.g., by defining their own \gls{ci-cd}
workflows based on provided templates. This way, our outputs also have a greater potential of becoming sustainable: once they are
persistently distributed, no continuous funding for HERMES is needed to use them. Instead, the community may apply further funding
as needed, e.g., for further development.

\subsubsection{Expectations on users and sources}\label{subsubsec:concept-scope-expect}
HERMES cannot clean up messy projects for users. Instead all tooling relies on the user to provide

\begin{itemize}  
    \item well-structured source code repositories
    \item with separated artifacts for data and software and
    \item with any possible legal issues resolved beforehand and appropriatly chosen licenses.
\end{itemize}

Additionally, we rely on users to supply any authentication credentials needed for workflows to run successfully. 
Examples include target publication repositories APIs source code platform APIs, continuous integration systems and the like. 

On a side note: usage scenarios with multiple metadata sources are likely to be the norm, not the exception. Please see \crtlnameref{subsec:concept-metadata-sources} for details.

\subsubsection{In scope}\label{subsubsec:concept-scope-in}
The user receives assistance in depositing software in an automated fashion. This may be used to create publications purely with 
rich metadata (to be at least FAIR~\cite{FAIR4RS-Principles}, even for closed source software) or with attached artifacts like 
source code, executables, etc. (to be more easily reusable). To achieve this, HERMES provides

\begin{figure}[htp]
    \centering
    \includegraphics[height=0.85\textheight]{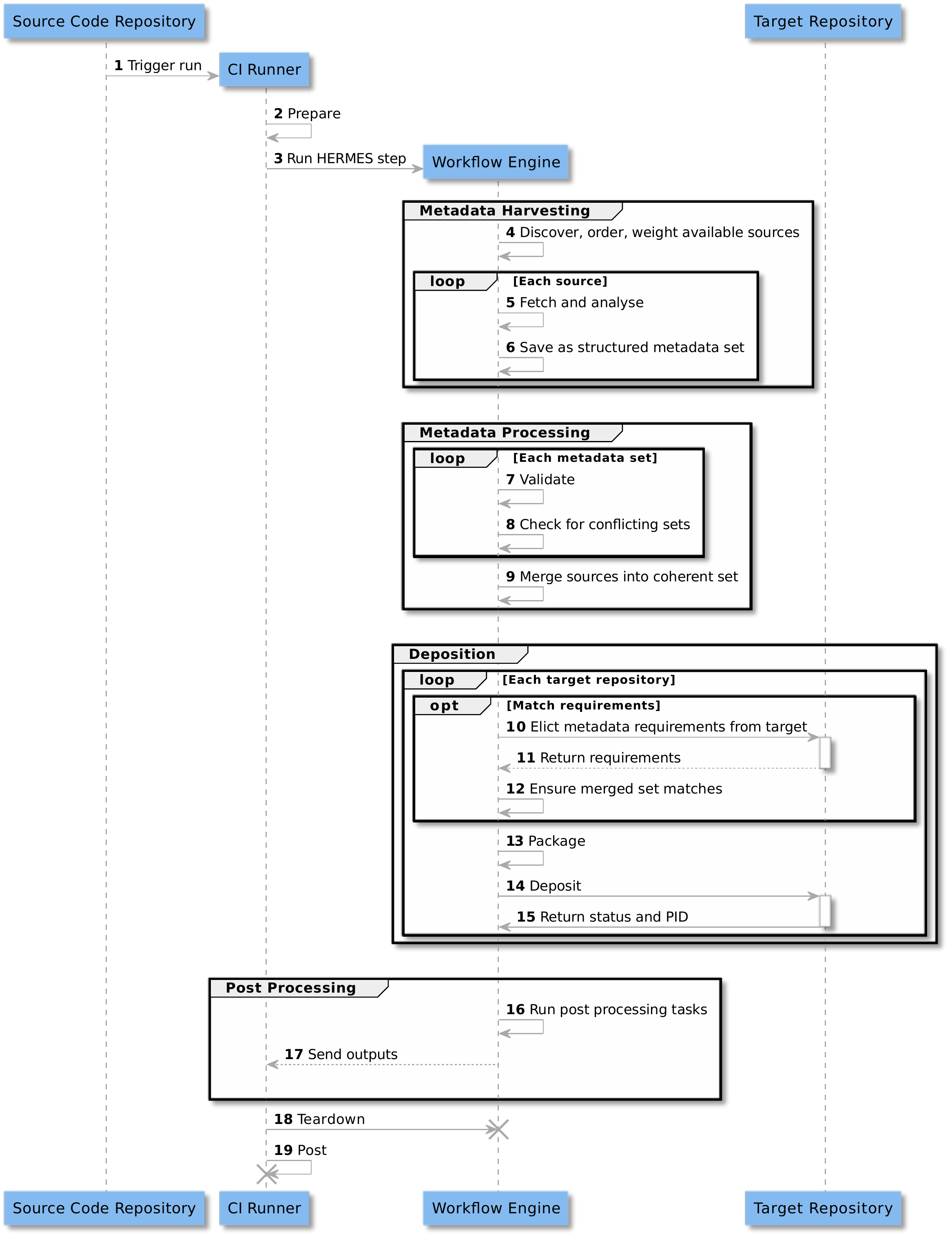}
    
    \caption{Sequence diagram of HERMES workflow pipelines}
    \caption*{Simple use case (1 software, n targets) data flow, showing how HERMES workflow pipelines kick into action.
              Outputs of any pipeline are forwarded to the next pipeline. Both \gls{ci-cd} configuration and pipelines
              allow for easy modifications or extensions if necessary.}
    \label{fig:hermes-workflow}
\end{figure}

\begin{itemize}  
    \item an extensible, configurable and automatable toolchain with capability to 
          be executed for\footnote{Please take a look at figure \ref{fig:hermes-workflow} for a more visual explanation.}
        \begin{itemize}  
            \item N software publications in 
            \item M target publication repositories 
            \item from the same origin
            \item as configured by the user,
        \end{itemize}
    \item initially harvesting and collating \gls{static-metadata} from formerly described \crtlnameref{subsec:concept-metadata-sources} and
    \item initially targeting
        \begin{itemize}  
            \item \nameref{par:repos-inveniordm} and
            \item \nameref{par:repos-dataverse}
        \end{itemize}
    \item for deposits of metadata and artifacts according to curator-defined requirements
    \item and output of the respective metadata in a structured format (e.g., \nameref{par:metadata-formats-codemeta}) for further reuse.
\end{itemize}

\subsubsection{Future scopes and extensions}\label{subsubsec:concept-scope-future}
In the future HERMES’ extensible architecture can be used to also support archives, such as Software Heritage~\cite{SWHArchive}, 
and domain and other registries such as ASCL~\cite{ASCL-Software-Citation}, the Research Software Directory~\cite{RSD-2018}, etc.

\subsection{Implementation outline}\label{subsec:concept-implementation}
As discussed in \nameref{subsubsec:concept-scope-in}, we provide metadata tooling and templates to integrate this tooling in
automated (\gls{ci-cd}) workflows. In this section, we briefly describe the basic concepts for the implementation of this tooling.

\subsubsection{Architecture}\label{subsubsec:concept-implementation-architecture}
\begin{figure}[htp]
    \centering
    \includegraphics[height=0.9\textheight]{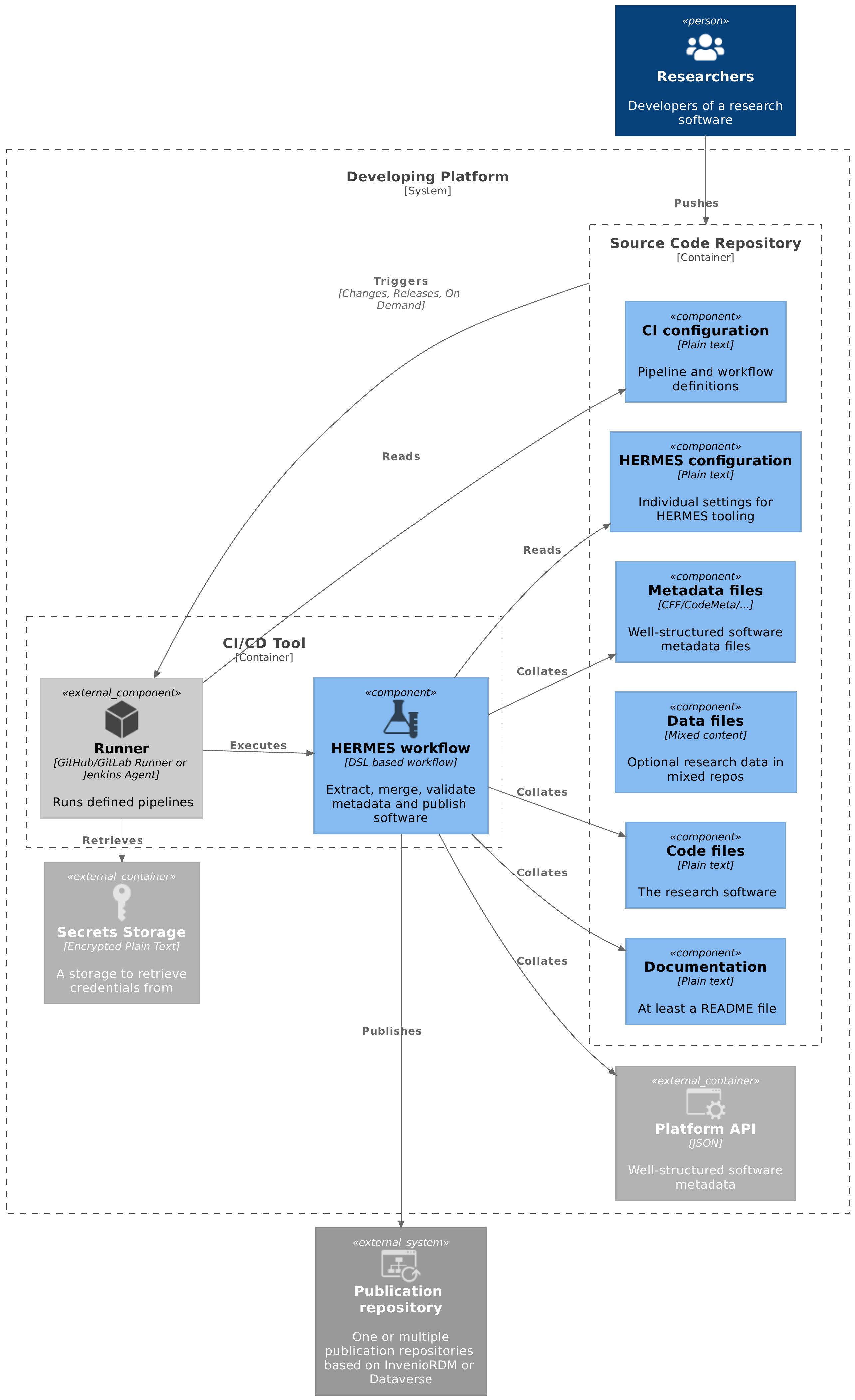}
    
    \caption{C4 Component architecture diagram}
    \caption*{Outlining how the HERMES workflow relates to researchers, gets embedded in runners, collates metadata from sources and publishes in target publication repositories.}
    \label{fig:hermes-architecture}
\end{figure}

As described within \crtlnameref{subsec:concept-scope}, our implementation reuses existing computing resources by leveraging workflows
on them. Figure \ref{fig:hermes-architecture} on page \pageref{fig:hermes-architecture} uses a \href{https://c4model.com/#ComponentDiagram}{C4 component diagram}
to outline the overall architecture of our solution.

\subsubsection{Workflow pipeline modeling}\label{subsubsec:concept-implementation-workflow}
As figure \ref{fig:hermes-workflow} on page \pageref{fig:hermes-workflow} outlines, HERMES implements four discrete pipelines with
public interfaces for extensibility and based on existing state-of-the-art \crtlnameref{sec:tooling} where possible, namely

\begin{enumerate}  
    \item a metadata harvesting pipeline that
        \begin{itemize}  
            \item runs a metadata analysis to determine the concrete harvesting tools to apply to the discovered \crtlnameref{subsec:concept-metadata-sources}, and
            \item retrieves the existing metadata from them;
        \end{itemize}
    \item a metadata processing pipeline that
        \begin{itemize}  
            \item validates the retrieved metadata, i.e., checks for conflicting sources, and
            \item merges them into a coherent set;
        \end{itemize}
    \item a metadata deposition pipeline that
        \begin{itemize}  
            \item optionally elicits metadata requirements from target \crtlnameref{subsubsec:repositories} and matches the merged set against them,
            \item publishes the set of metadata with or without the respective software artifacts to those target repositories, and
            \item retrieves the persistent identifier for the deposition; and
        \end{itemize}
    \item a post-processing pipeline that 
        \begin{itemize}  
            \item optionally updates metadata in the source repositories, e.g., with the deposition identifier,
            \item notifies users of any issues that were encountered during the workflow run, and
            \item passes the software and deposit metadata to any following steps in the users’ \gls{ci-cd} workflow.
        \end{itemize}
\end{enumerate}

We also provide reference implementations for commonly used continuous integration tools, such as GitLab CI, GitHub Actions
and Jenkins that combine the four pipelines into a complete solution for automated publication of software with rich metadata
(see also \ref{subsubsec:concept-implementation-architecture}).

\subsubsection{Adding missing functionalities to target repository software}\label{subsubsec:concept-implementation-targets}
To enable the metadata deposition pipeline described above to publish software with rich metadata, the two targeted publication
repositories need to be prepared to accept metadata sets compiled in the metadata processing pipeline.

Both the \nameref{par:repos-dataverse} and \nameref{par:repos-inveniordm} lack some features for advanced
software metadata intake and presentation. The current iteration of the HERMES project investigates and coordinates with these
projects and stakeholders to add any missing functionalities upstream.

\subsubsection{Templates, documentation and training resources}\label{subsubsec:templates-documentation-training}

HERMES will safeguard the usability and sustainability of the implemented tooling by enabling the growth of a community through three main strategies: exemplary workflow templates, comprehensive documentation and the provision of training resources for end users.

We provide exemplary workflows for combinations of commonly used \gls{ci-cd} systems and our target \gls{pubrepo}. 
These templates will be published under open licenses and can be adapted by end users to suit their needs.

The documentation that the project produces encompasses conceptual documentation for stakeholders (of which this paper
is a starting point), technical documentation for future developers and maintainers as well as integrating parties,
and documentation for end users, i.e., researchers looking to publish their software with HERMES tooling.

Furthermore, we develop training resources for end users.
These resources are planned to be implemented in training curricula within the Helmholtz Association as part of the HIFIS project.
Additionally, they are being made available for reuse by the wider community, and licensed under open licenses.

\clearpage

\appendix
\addsec{Tables and Figures}\label{tables-figures}


\begin{table}[ht]
    \newcommand*\rot[1]{\hbox to1em{\hss\rotatebox[origin=br]{-60}{#1}}}
    \newcommand*\C[1]{\ifcase#1 -\or\LEFTcircle\or\CIRCLE\fi}
    
    \centering
    
    \begin{threeparttable}
        \begin{tabular}{r *{8}{c} @{\hskip 10mm}*{2}{c} @{\hskip 10mm}c}
            \toprule
            Metadata type & \multicolumn{11}{c}{Metadata format}\\
            \midrule
             & \rot{Plain Text files}
             & \rot{CodeMeta files}
             & \rot{Citation File Format}
             & \rot{Zenodo JSON files}
             & \rot{BibTex files}
             & \rot{Manifest files}
             & \rot{Configuration files}
             & \rot{Linked Data files}
             & \rot{Version control sys.}
             & \rot{Platform APIs}
             & \rot{Other}
             \\
            \midrule
            Software name                   &\C2&\C2&\C2&\C2&\C2&\C2&\C0&\C2&\C1&\C2&\C0 \\
            File system metadata            &\C0&\C0&\C0&\C0&\C0&\C0&\C0&\C2&\C1&\C0&\C2 \\
            Authorship information          &\C2&\C2&\C2&\C2&\C2&\C1&\C1&\C2&\C1&\C1&\C0 \\
            Documentation reference         &\C2&\C2&\C0&\C0&\C0&\C0&\C2&\C2&\C0&\C2&\C2 \\
            Legal and licensing info.       &\C2&\C2&\C2&\C2&\C0&\C2&\C0&\C2&\C0&\C2&\C2 \\
            Funding information             &\C2&\C2&\C0&\C2&\C0&\C0&\C0&\C2&\C0&\C0&\C2 \\
            Domain context                  &\C2&\C0&\C0&\C0&\C0&\C1&\C0&\C2&\C0&\C1&\C2 \\
            Citation metrics                &\C2&\C0&\C0&\C0&\C0&\C0&\C0&\C2&\C0&\C2&\C2 \\
            Location metadata               &\C2&\C2&\C2&\C0&\C1&\C0&\C1&\C2&\C2&\C2&\C2 \\
            Publication dates, etc.         &\C0&\C2&\C2&\C2&\C2&\C0&\C0&\C2&\C1&\C2&\C0 \\
            Categorization inform.          &\C2&\C2&\C2&\C2&\C0&\C1&\C0&\C2&\C0&\C1&\C2 \\
            Availability information        &\C2&\C0&\C0&\C0&\C0&\C0&\C0&\C2&\C0&\C0&\C2 \\
            Identifiers                     &\C2&\C2&\C2&\C2&\C2&\C0&\C0&\C2&\C1&\C1&\C2 \\
            Relational metadata             &\C2&\C0&\C0&\C2&\C0&\C2&\C0&\C2&\C1&\C1&\C0 \\
            High-level description          &\C2&\C2&\C2&\C2&\C1&\C1&\C0&\C2&\C0&\C1&\C0 \\
            Dependency information          &\C2&\C2&\C2&\C0&\C0&\C2&\C1&\C2&\C0&\C1&\C0 \\
            Lines of code                   &\C0&\C0&\C0&\C0&\C0&\C0&\C0&\C2&\C1&\C2&\C2 \\
            Programming language            &\C2&\C0&\C0&\C0&\C0&\C2&\C2&\C2&\C0&\C2&\C0 \\
            Version information             &\C2&\C2&\C2&\C2&\C1&\C2&\C1&\C2&\C2&\C2&\C2 \\
            Runtime requirements            &\C2&\C0&\C0&\C0&\C0&\C2&\C2&\C2&\C0&\C0&\C2 \\
            References                      &\C2&\C2&\C2&\C2&\C0&\C2&\C1&\C2&\C0&\C2&\C2 \\
            Software quality metrics        &\C2&\C0&\C0&\C0&\C0&\C0&\C2&\C2&\C0&\C2&\C2 \\
            Development metrics             &\C2&\C0&\C0&\C0&\C0&\C0&\C0&\C2&\C2&\C2&\C2 \\
            Usage metrics                   &\C2&\C0&\C0&\C0&\C0&\C0&\C0&\C2&\C0&\C2&\C2 \\
            Infrastructural metadata        &\C2&\C0&\C0&\C0&\C0&\C0&\C2&\C2&\C1&\C2&\C2 \\
            \bottomrule
        \end{tabular}
        \begin{tablenotes}
            \item $\C2=\text{provides type}$; $\C1=\text{partially prov. type}$; $\text{\C0}=\text{does not prov. type}$;
        \end{tablenotes}
    \end{threeparttable}
    
    \caption{Mapping metadata types to common metadata formats.}
    \label{tab:comparision-metadata-types}
\end{table}

\clearpage

\begin{table}[htp]
    \newcommand*\rot[1]{\hbox to1em{\hss\rotatebox[origin=br]{-45}{#1}}}
    \newcommand*\webhook{\faLink}
    \newcommand*\harvest{\faTractor}
    \newcommand*\cihook{\faCloudUpload*}
    \newcommand*\service{\faConciergeBell}
    \newcommand*\scripts{\faScroll}
    \newcommand*\file{\faFile*}
    \newcommand*\vcs{\faCodeBranch}
    \newcommand*\api{\faCloudDownload*}
    \newcommand*\yes{\faCheck}
    \newcommand*\kinda{(\yes)}
    \newcommand*\unknown{?}
    \newcommand*\no{-}
    \newcommand*\rs{\repostatus}
    \renewcommand*\sup[1]{$^\textrm{#1}$}
    
    \newcommand*\citeRSD{\cite{sw:rsd}}
    \newcommand*\citeHZDR{\cite{sw:rodare-bridge}}
    \newcommand*\citeAMES{\cite{ASCL-CaltechAMES}}
    \newcommand*\citeSARA{\sup{i},\cite{SARA}}
    \newcommand*\citeCARP{\cite{sw:opencarp-ci}}
    \newcommand*\citeZenodraft{\cite{sw:zenodraft},\cite{sw:zenodraft-action}}

    \centering
    
    \begin{threeparttable}
        \begin{tabular}{r *7{c} @{\hskip 10mm} *4{c}}
            \toprule
                & \multicolumn{7}{l}{Pull-based}
                & \multicolumn{4}{l}{Push-based}\\
            \midrule
                & \rot{Research Software Directory}
                & \rot{Software Heritage Archive}
                & \rot{Sw. Heritage Github Action (\ref{par:workflows-swh-gh-action})}
                & \rot{CaltechDATA AMES}
                & \rot{GitHub-Zenodo Integration}
                & \rot{GitLab-Zenodo Feature Request}
                & \rot{GitLab-InvenioRDM Integration}
                & \rot{Zenodraft Github Action (\ref{par:workflows-zenodraft})}
                & \rot{OpenCARP CI}
                & \rot{SARA service}
                & \rot{Preservation Quality Tool}\\
            \midrule
                Status$^\star$     & \rs{Active} & \rs{Active} & \rs{Suspended} & \rs{Active} & \rs{Active}   & \rs{Concept} & \rs{Suspended} & \rs{Suspended} & \rs{Active} & \rs{Abandoned} & \rs{Unsupported} \\
                Type               & \harvest    & \harvest    & \cihook        & \webhook    & \webhook      & \webhook     & \webhook       & \scripts       & \scripts    & \service       & \service         \\
                Documentation      & \citeRSD    & \sup{a,b}   & \sup{c}        & \citeAMES   & \sup{d,e}     & \sup{g,h}    & \citeHZDR      & \citeZenodraft & \citeCARP   & \citeSARA      & \sup{j}          \\
            \midrule
            Extract metadata from: & \yes        & \yes        & \kinda         & \yes        & \yes          & \unknown     & \yes           & \yes           & \yes        & \yes           & \yes             \\
                 Zenodo JSON \file & \no         & \no         & \no            & \no         & \yes\sup{f}   & \unknown     & \yes           & \yes           & \no         & \no            & \no              \\
                    CodeMeta \file & \no         & \yes        & \kinda         & \yes        & \no           & \unknown     & \no            & \no            & \no         & \no            & \no              \\
        Citation File Format \file & \yes        & \yes        & \kinda         & \no         & \yes\sup{f}   & \unknown     & \no            & \no            & \no         & \no            & \no              \\
        Other via Crosswalks \file & \no         & \yes        & \kinda         & \no         & \no           & \unknown     & \no            & \no            & \no         & \no            & \no              \\
                   Plaintext \file & \no         & \no         & \no            & \no         & \no           & \unknown     & \no            & \no            & \no         & \yes           & \no              \\
               Configuration \file & \no         & \no         & \no            & \no         & \no           & \unknown     & \no            & \no            & \yes        & \no            & \yes             \\
       Version control system \vcs & \no         & \no         & \no            & \no         & \no           & \unknown     & \no            & \no            & \yes        & \yes           & \no              \\
           Platform API resp. \api & \yes        & \no         & \no            & \no         & \no           & \unknown     & \no            & \no            & \no         & \no            & \yes             \\
            \midrule
        Create publication w/ m.d. & \yes        & \yes        & \kinda         & \no         & \yes          & \unknown     & \yes           & \yes           & \yes        & \yes           & \yes             \\
        Mint persistent identifier & \no         & \yes        & \kinda         & \no         & \yes          & \unknown     & \yes           & \yes           & \yes        & \yes           & \yes             \\
          Update existing metadata & \yes        & \no         & \no            & \yes        & \no           & \unknown     & \no            & \yes           & \no         & \no            & \no              \\
                in Zenodo          & \no         & \no         & \no            & \no         & \yes          & \unknown     & \yes           & \yes           & \no         & \no            & \yes             \\
                in InvenioRDM      & \no         & \no         & \no            & \yes        & \kinda\sup{e} & \unknown     & \yes           & \no            & \no         & \no            & \no              \\
                in Sw. Heritage    & \no         & \yes        & \kinda         & \no         & \no           & \no          & \no            & \no            & \no         & \no            & \no              \\
                in Other           & \yes        & \no         & \no            & \no         & \no           & \no          & \no            & \no            & \yes        & \yes           & \yes             \\
            \bottomrule
        \end{tabular}
        \begin{tablenotes}
            \footnotesize
            \item Note: depositing software artifacts may be part of some of these workflows. \\ To keep the table focused on complexer metadata issues, this is left out on purpose.
            \item
            \item Status$^\star$: \repostatus[label]{Active}, \repostatus[label]{Suspended}, \repostatus[label]{Concept}, \repostatus[label]{Abandoned}, \repostatus[label]{Unsupported}
            \item Type: \webhook\,Webhook, \harvest\,Harvesting, \cihook\,\gls{ci-cd}-based Webhook, \scripts\,Script based, \service\,Web service
            \item Support: \yes\,supported, \kinda\,indirectly supported, \unknown\,not yet known, \no\,unsupported
            \item Sources: \file\,file-based
            \item
            \item[$\star$]  Using controlled vocabulary \gls{repostatus}
            \item[a]        See \url{https://archive.softwareheritage.org/save} and \cite{SWHArchive}
            \item[b]        See \href{https://docs.softwareheritage.org/devel/swh-indexer/metadata-workflow.html}{SWH Indexer Metadata Workflow}, referenced from \cite{FennerBlogCff}
            \item[c]        See section \nameref{par:workflows-swh-gh-action} (\ref{par:workflows-swh-gh-action})
            \item[d]        \url{https://guides.github.com/activities/citable-code} or \\ \url{https://developers.zenodo.org/\#add-metadata-to-your-github-repository-release}
            \item[e]        Zenodo extends \url{https://github.com/inveniosoftware/invenio-github} (CFF since \href{https://github.com/inveniosoftware/invenio-github/pull/89}{PR 89})
            \item[f]        Zenodo JSON and CFF mutually exclusive. See \href{https://twitter.com/ZENODO_ORG/status/1420357001490706442}{announcement},
                            \href{https://github.com/zenodo/zenodo/pull/2201}{Zenodo PR} \& \href{https://github.com/zenodo/zenodo/commit/a806d10700a504cc86abaf9eb961dc201b3d8fe8}{Zenodo Fix}
            \item[g]        \url{https://github.com/zenodo/zenodo/issues/1404}
            \item[h]        \url{https://gitlab.com/gitlab-org/gitlab/-/issues/25587}
            \item[i]        \url{https://sara-service.org}        
            \item[j]        \url{https://presqt.readthedocs.io}
        \end{tablenotes}
    \end{threeparttable}
    
    \caption{Existing software publication workflows and building bricks}
    \label{tab:workflows}
\end{table}

\clearpage

\clearpage
\addsec{Acknowledgments}\label{acknowledgments}
This project (ZT-I-PF-3-006) was funded by the \enquote{\href{https://www.helmholtz.de/en/about-us/structure-and-governance/initiating-and-networking/}{Initiating and Networking Fund of the Helm\-holtz Association}}
in the framework of the \enquote{\href{https://helmholtz-metadaten.de/en}{Helmholtz Metadata Collaboration}} project call.

We thank the participants of the \href{https://events.hifis.net/event/205/}{project kickoff workshop} for their contributions to the project plan as well as their comments to this document. We especially thank Daniel Garijo (Universidad Politécnica de Madrid), Carlos Martinez-Ortiz (Netherlands eScience Center), Ana Trisovic (Harvard University), Sara Gonzalez (Northwestern University), Dorothea Iglezakis (University Library Stuttgart) and Felix Bach (Karlsruhe Institute of Technology) for presenting their work, as well as Deborah Schmidt (MDC Berlin), Ronny Gey (UFZ Leipzig), Anton Pirogov (FZ Jülich), Dennis Gläser (University of Stuttgart), Jens Bröder (FZ Jülich), Oliver Karras (TIB), Pedro Videgain Barranco (FZ Jülich), Anett Seeland (University of Stuttgart), Kirsten Elger (GFZ German Research Centre for Geosciences) and Jan Göpfert (FZ Jülich).

We highly appreciate the conducted draft paper reviews by Uwe Konrad (HZDR), Bernhard Mittermaier (FZ Jülich), Carina Haupt (DLR) and Ana Trisovic (Harvard University).

We also thank Fonticons, Inc. for free usage of the \href{https://fontawesome.com}{FontAwesome icons} under a \href{https://fontawesome.com/license/free}{CC-BY license} throughout this document.

\setglossarystyle{altlist}
\printglossaries

\phantomsection
\printbibliography[heading=bibintoc]

@misc{ASCL-Software-Citation,
      title={Citation method, please? A case study in astrophysics}, 
      author={Alice Allen},
      year={2021},
      eprint={2111.12574},
      archivePrefix={arXiv},
      primaryClass={astro-ph.IM}
}

@online{ASCL-CaltechAMES,
  title        = {Software in the CaltechDATA Repository},
  author       = {Morrell, Thomas E.},
  year         = 2019,
  month        = {11},
  url          = {https://github.com/ASCLnet/SWRegistryWorkshop/blob/2cd02ad1862ebc67ad9fcb0347b832d10d8c7ca9/presentations/Caltech-Software-Presentation.pptx},
  place        = {University of Maryland}
}

@book{PeerJSoftVsDataCitation,
  title        = {Software vs. data in the context of citation},
  author       = {Katz, Daniel S. and Niemeyer, Kyle E. and Smith, Arfon M. and Anderson, William L. and Boettiger, Carl and Hinsen, Konrad and Hooft, Rob and Hucka, Michael and Lee, Allen and Löffler, Frank and Pollard, Tom and Rios, Fernando},
  year         = 2016,
  month        = {12},
  number       = {e2630v1},
  doi          = {10.7287/peerj.preprints.2630v1},
  url          = {https://peerj.com/preprints/2630},
  abstractnote = {Software is data, but it is not just data. While “data” in computing and information science can refer to anything that can be processed by a computer, software is a special kind of data that can be a creative, executable tool that operates on data. However, software and data are similar in that they both traditionally have not been cited in publications. This paper discusses the differences between software and data in the context of citation, by providing examples and referring to evidence in the form of citations.},
  institution  = {PeerJ Inc.}
}

@article{TrackingCitations,
  title        = {Practice meets Principle: Tracking Software and Data Citations to Zenodo DOIs},
  author       = {van de Sandt, Stephanie and Nielsen, Lars Holm and Ioannidis, Alexandros and Muench, August and Henneken, Edwin and Accomazzi, Alberto and Bigarella, Chiara and Lopez, Jose Benito Gonzalez and Dallmeier-Tiessen, Sünje},
  year         = 2019,
  month        = {11},
  journal      = {arXiv:1911.00295 [cs]},
  url          = {http://arxiv.org/abs/1911.00295},
  note         = {arXiv: 1911.00295},
  abstractnote = {Data and software citations are crucial for the transparency of research results and for the transmission of credit. But they are hard to track, because of the absence of a common citation standard. As a consequence, the FORCE11 recently proposed data and software citation principles as guidance for authors. Zenodo is recognized for the implementation of DOIs for software on a large scale. The minting of complementary DOIs for the version and concept allows measuring the impact of dynamic software. This article investigates characteristics of 5,456 citations to Zenodo data and software that were captured by the Asclepias Broker in January 2019. We analyzed the current state of data and software citation practices and the quality of software citation recommendations with regard to the impact of recent standardization efforts. Our findings prove that current citation practices and recommendations do not match proposed citation standards. We consequently suggest practical first steps towards the implementation of the software citation principles.}
}

@article{SoftCitePrinciples,
  title        = {Software citation principles},
  author       = {Smith, Arfon M. and Katz, Daniel S. and Niemeyer, Kyle E. and FORCE11 Software Citation Working Group},
  year         = 2016,
  journal      = {PeerJ Computer Science},
  volume       = 2,
  number       = {e86},
  doi          = {10.7717/peerj-cs.86},
  url          = {https://doi.org/10.7717/peerj-cs.86}
}

@misc{FAIR4RS-FreshLook,
  title={A Fresh Look at FAIR for Research Software}, 
  author={Daniel S. Katz and Morane Gruenpeter and Tom Honeyman and Lorraine Hwang and Mark D. Wilkinson and Vanessa Sochat and Hartwig Anzt and Carole Goble and for FAIR4RS Subgroup 1},
  year={2021},
  month        = {2},
  eprint={2101.10883},
  archivePrefix={arXiv},
  primaryClass={cs.SE},
  abstractnote = {This document captures the discussion and deliberation of the FAIR for Research Software (FAIR4RS) subgroup that took a fresh look at the applicability of the FAIR Guiding Principles for scientific data management and stewardship for research software. We discuss the vision of research software as ideally reproducible, open, usable, recognized, sustained and robust, and then review both the characteristic and practiced differences of research software and data. This vision and understanding of initial conditions serves as a backdrop for an attempt at translating and interpreting the guiding principles to more fully align with research software. We have found that many of the principles remained relatively intact as written, as long as considerable interpretation was provided. This was particularly the case for the “Findable” and “Accessible” foundational principles. We found that “Interoperability” and “Reusability” are particularly prone to a broad and sometimes opposing set of interpretations as written. We propose two new principles modeled on existing ones, and provide modified guiding text for these principles to help clarify our final interpretation. A series of gaps in translation were captured during this process, and these remain to be addressed. We finish with a consideration of where these translated principles fall short of the vision laid out in the opening.}
}

@unpublished{FAIR4RS-Principles,
  title        = {FAIR Principles for Research Software (FAIR4RS Principles)},
  author       = {Chue Hong, Neil P. and Katz, Daniel S. and Barker, Michelle and Lamprecht, Anna-Lena and Martinez, Carlos and Psomopoulos, Fotis E. and Harrow, Jen and Castro, Leyla Jael and Gruenpeter, Morane and Martinez, Paula Andrea and Honeyman, Tom and Struck, Alexander and Lee, Allen and Loewe, Axel and van Werkhove, Ben and Jones, Catherine and Garijo, Daniel and Plomp, Esther and Genova, Francoise and Shanahan, Hugh and Leng, Joanna and Hellström, Maggie and Sandström, Malin and Sinha, Manodeep and Kuzak, Mateusz and Herterich, Patricia and Zhang, Qian and Islam, Sharif and Sansone, Susanna-Assunta and Pollard, Tom and Atmojo, Udayanto Dwi and Williams, Alan and Czerniak, Andreas and Niehues, Anna and Fouilloux, Anne Claire and Desinghu, Bala and Goble, Carole and Richard, Céline and Gray, Charles and Erdmann, Chris and Nüst, Daniel and Tartarini, Daniele and Ranguelova, Elena and Anzt, Hartwig and Todorov, Ilian and McNally, James and Moldon, Javier and Burnett, Jessica and Garrido-Sánchez, Julián and Belhajjame, Khalid and Sesink, Laurents and Hwang, Lorraine and Tovani-Palone, Marcos Roberto and Wilkinson, Mark D. and Servillat, Mathieu and Liffers, Matthias and Fox, Merc and Miljković, Nadica and Lynch, Nick and Martinez Lavanchy, Paula and Gesing, Sandra and Stevens, Sarah and Martinez Cuesta, Sergio and Peroni, Silvio and Soiland-Reyes, Stian and Bakker, Tom and Rabemanantsoa, Tovo and Sochat, Vanessa and Yehudi, Yo},
  year         = 2021,
  doi          = {10.15497/RDA00065},
  url          = {https://doi.org/10.15497/RDA00065},
  note         = {Publisher: Research Data Alliance},
  abstractnote = {Research software is a fundamental and vital part of research worldwide, yet there remain significant challenges to software productivity, quality, reproducibility, and sustainability. Improving the practice of scholarship is a common goal of the open science, open source software and FAIR (Findable, Accessible, Interoperable and Reusable) communities, but improving the sharing of research software has not yet been a strong focus of the latter. To improve the FAIRness of research software, the FAIR for Research Software (FAIR4RS) Working Group has sought to understand how to apply the FAIR Guiding Principles for scientific data management and stewardship to research software, bringing together existing and new community efforts. Many of the FAIR Guiding Principles can be directly applied to research software by treating software and data as similar digital research objects. However, specific characteristics of software — such as its executability, composite nature, and continuous evolution and versioning — make it necessary to revise and extend the principles. This document presents the first version of the FAIR Principles for Research Software (FAIR4RS Principles). It is an outcome of the FAIR for Research Software Working Group (FAIR4RS WG). The FAIR for Research Software Working Group is jointly convened as an RDA Working Group, FORCE11 Working Group, and Research Software Alliance (ReSA) Task Force.}
}

@article{DFG-GRP19,
  title        = {Guidelines for Safeguarding Good Research Practice. Code of Conduct},
  author       = {Forschungsgemeinschaft, Deutsche},
  year         = 2019,
  month        = {9},
  doi          = {10.5281/zenodo.3923602},
  url          = {https://www.dfg.de/download/pdf/foerderung/rechtliche_rahmenbedingungen/gute_wissenschaftliche_praxis/kodex_gwp_en.pdf},
  abstractnote = {The DFG´s Code of Conduct “Safeguarding Good Research Practice” represents the consensus among the member organisations of the DFG on the fundamental principles and standards of good practice and are upheld by these organisations. These guidelines underline the importance of integrity in the everyday practice of research and provide researchers with a reliable reference with which to embed good research practice as an established and binding aspect of their work.}
}

@article{HelmholtzSWPolicy,
  title        = {Muster-Richtlinie Nachhaltige Forschungssoftware an den Helmholtz-Zentren},
  author       = {Task Group Forschungssoftware and Bach, Felix and Bertuch, Oliver and Busse, Christian and Castell, Wolfgang zu and Celo, Sabine and Denker, Michael and Dinkelacker, Stefan and Druskat, Stephan and Faber, Claas and Finke, Ants and Fritzsch, Bernadette and Hammitzsch, M. and Haseleu, Julia and Konrad, Uwe and Krupa, Jörn and Leifels, Yvonne and Mohns-Pöschke, Kerstin and Moravcikova, Martina and Nöller, Joachim and Möhl, Christoph and Nolden, Marco and Scheinert, Markus and Schelhaas, Ute and Scheliga, Katharina Sara and Schlauch, Tobias and Schnicke, Thomas and Scholz, Almut and Schwennsen, Florian and Seifarth, Jenny and Selzer, Michael and Shishatskiy, Sergey and Steglich, Dirk and Strohbach, Sandra and Terhorst, Dennis and Al-Turany, Mohammad and Vierkant, Paul and Wieser, Thomas and Witter, Ludwig and Wortmann, Daniel},
  year         = 2019,
  doi          = {10.2312/os.helmholtz.007},
  url          = {https://gfzpublic.gfz-potsdam.de/pubman/faces/ViewItemOverviewPage.jsp?itemId=item_4906899},
  note         = {Publisher: Helmholtz Open Science Office},
  abstractnote = {Autor: Task Group Forschungssoftware et al.; Genre: Sonstige; Final veröffentlicht: 2019; Open Access; Titel: Muster-Richtlinie Nachhaltige Forschungssoftware an den Helmholtz-Zentren}
}

@article{SWORD,
  title        = {SWORD: Simple Web-service Offering Repository Deposit},
  author       = {Allinson, Julie and Francois, Sebastien and Lewis, Stuart},
  year         = 2008,
  journal      = {Ariadne},
  number       = 54,
  url          = {http://www.ariadne.ac.uk/issue/54/allinson-et-al/}
}

@article{DataCiteSchema,
  title        = {DataCite Metadata Schema Documentation for the Publication and Citation of Research Data and Other Research Outputs v4.4},
  author       = {DataCite Metadata Working Group},
  year         = 2021,
  pages        = {82 pages},
  doi          = {10.14454/3W3Z-SA82},
  url          = {https://schema.datacite.org/meta/kernel-4.4/},
  note         = {Artwork Size: 82 pages Medium: application/pdf Publisher: DataCite Version Number: 4.4}
}

@article{RO-Crate,
  title        = {Packaging research artefacts with RO-Crate},
  author       = {Soiland-Reyes, Stian and Sefton, Peter and Crosas, Mercè and Castro, Leyla Jael and Coppens, Frederik and Fernández, José M. and Garijo, Daniel and Grüning, Björn and La Rosa, Marco and Leo, Simone and Carragáin, Eoghan Ó and Portier, Marc and Trisovic, Ana and RO-Crate Community and Groth, Paul and Goble, Carole},
  year         = 2021,
  month        = {8},
  doi          = {10.5281/ZENODO.5146228},
  isbn         = 9781315351148,
  url          = {https://zenodo.org/record/5146228},
  note         = {Publisher: Zenodo},
  abstractnote = {An increasing number of researchers support reproducibility by including pointers to and descriptions of datasets, software and methods in their publications. However, scientific articles may be ambiguous, incomplete and difficult to process by automated systems. In this paper we introduce RO-Crate, an open, community-driven, and lightweight approach to packaging research artefacts along with their metadata in a machine readable manner. RO-Crate is based on Schema.org annotations in JSON-LD, aiming to establish best practices to formally describe metadata in an accessible and practical way for their use in a wide variety of situations. An RO-Crate is a structured archive of all the items that contributed to a research outcome, including their identifiers, provenance, relations and annotations. As a general purpose packaging approach for data and their metadata, RO-Crate is used across multiple areas, including bioinformatics, digital humanities and regulatory sciences. By applying “just enough” Linked Data standards, RO-Crate simplifies the process of making research outputs FAIR while also contributes to enhancing research reproducibility. An RO-Crate for this article is available.}
}

@book{BagItRFC,
  title        = {The BagIt File Packaging Format (V1.0)},
  author       = {Kunze, J. and Littman, J. and Madden, E. and Scancella, J. and Adams, C.},
  year         = 2018,
  number       = {RFC8493},
  pages        = {RFC8493},
  doi          = {10.17487/RFC8493},
  url          = {https://www.rfc-editor.org/info/rfc8493},
  institution  = {RFC Editor}
}

@misc{W3C-RDF,
  title        = {RDF 1.1 Primer},
  url          = {https://www.w3.org/TR/rdf11-primer/}
}

@misc{W3C-JSON-LD,
  title        = {W3C JSON-LD},
  url          = {https://www.w3.org/TR/json-ld/}
}

@misc{W3C-RDF-Turtle,
  title        = {W3C RDF 1.1 Turtle},
  url          = {https://www.w3.org/TR/turtle/}
}

@misc{W3C-SPARQL,
  title        = {SPARQL Query Language for RDF},
  url          = {https://www.w3.org/TR/rdf-sparql-query/}
}

@article{SoftwareOntology,
  title        = {The Software Ontology (SWO): a resource for reproducibility in biomedical data analysis, curation and digital preservation},
  author       = {Malone, James and Brown, Andy and Lister, Allyson L. and Ison, Jon and Hull, Duncan and Parkinson, Helen and Stevens, Robert},
  year         = 2014,
  month        = {6},
  journal      = {Journal of Biomedical Semantics},
  volume       = 5,
  number       = 1,
  pages        = 25,
  doi          = {10.1186/2041-1480-5-25},
  url          = {https://doi.org/10.1186/2041-1480-5-25},
  abstractnote = {Biomedical ontologists to date have concentrated on ontological descriptions of biomedical entities such as gene products and their attributes, phenotypes and so on. Recently, effort has diversified to descriptions of the laboratory investigations by which these entities were produced. However, much biological insight is gained from the analysis of the data produced from these investigations, and there is a lack of adequate descriptions of the wide range of software that are central to bioinformatics. We need to describe how data are analyzed for discovery, audit trails, provenance and reproducibility.}
}

@misc{SoftDescOnto,
  title        = {The Software Description Ontology},
  author       = {Garijo, Daniel and Ratnakar, Varun and Gil, Yolanda and Khider, Deborah},
  year         = 2021,
  month        = {5},
  journal      = {The Software Description Ontology},
  url          = {https://w3id.org/okn/o/sd},
  abstractnote = {An ontology for describing software components, including their metadata (attribution, licensing, usage instructions, how to get support) and their inputs, outputs and variables. The ontology extends schema.org and CodeMeta vocabularies and is based on OntoSoft, which proposed a vocabulary for describing software by asking scientists questions.}
}

@book{DoapOntology,
  title        = {DOAP: Description Of A Project},
  author       = {Wilder-James, Edd},
  year         = 2021,
  month        = {11},
  url          = {https://github.com/ewilderj/doap},
  abstractnote = {RDF schema for describing software projects}
}

@article{SeonOntology,
  title        = {SEON: a pyramid of ontologies for software evolution and its applications},
  author       = {Würsch, Michael and Ghezzi, Giacomo and Hert, Matthias and Reif, Gerald and Gall, Harald C.},
  year         = 2012,
  month        = {11},
  journal      = {Computing},
  volume       = 94,
  number       = 11,
  pages        = {857–885},
  doi          = {10.1007/s00607-012-0204-1},
  url          = {https://doi.org/10.1007/s00607-012-0204-1},
  abstractnote = {The Semantic Web provides a standardized, well-established frameworkto define and work with ontologies. It is especially apt for machineprocessing. However, researchers in the field of software evolutionhave not really taken advantage of that so far. In this paper, weaddress the potential of representing software evolution knowledgewith ontologies and Semantic Web technology, such as Linked Data andautomated reasoning. We present Seon, a pyramid of ontologies forsoftware evolution, which describes stakeholders, their activities,artifacts they create, and the relations among all of them. We showthe use of evolution-specific ontologies for establishing a sharedtaxonomy of software analysis services, for defining extensiblemeta-models, for explicitly describing relationships amongartifacts, and for linking data such as code structures, issues(change requests), bugs, and basically any changes made to a systemover time. For validation, we discuss three different approaches,which are backed by Seon and enable semantically enriched softwareevolution analysis. These techniques have been fully implemented astools and cover software analysis with web services, a naturallanguage query interface for developers, and large-scale softwarevisualization.}
}

@article{SchemaOrg,
  title        = {Schema.org: evolution of structured data on the web},
  author       = {Guha, R. V. and Brickley, Dan and Macbeth, Steve},
  year         = 2016,
  month        = {1},
  journal      = {Communications of the ACM},
  volume       = 59,
  number       = 2,
  pages        = {44–51},
  doi          = {10.1145/2844544},
  url          = {https://dl.acm.org/doi/10.1145/2844544},
  abstractnote = {Big data makes common schemas even more necessary.}
}

@book{CodeMetaSchema,
  title        = {CodeMeta: an exchange schema for software metadata. Version 2.0},
  author       = {Jones, Matthew B. and Boettiger, Carl and Mayes, Abby Cabunoc and Smith, Arfon and Slaughter, Peter and Niemeyer, Kyle and Gil, Yolanda and Fenner, Martin and Nowak, Krzysztof and Hahnel, Mark and Coy, Luke and Allen, Alice and Crosas, Mercè and Sands, Ashley and Hong, Neil Chue and Cruse, Patricia and Katz, Dan and Goble, Carole},
  year         = 2017,
  doi          = {10.5063/schema/codemeta-2.0},
  url          = {https://doi.org/10.5063/schema/codemeta-2.0},
  note         = {Published: KNB Data Repository}
}

@inproceedings{SoMEF,
  title        = {SoMEF: A Framework for Capturing Scientific Software Metadata from its Documentation},
  author       = {Mao, A. and Garijo, D. and Fakhraei, S.},
  year         = 2019,
  booktitle    = {2019 IEEE International Conference on Big Data (Big Data)},
  pages        = {3032–3037},
  doi          = {10.1109/BigData47090.2019.9006447},
  url          = {http://dgarijo.com/papers/SoMEF.pdf}
}

@article{CffSchema,
  title        = {Citation File Format},
  author       = {Druskat, Stephan and Spaaks, Jurriaan H. and Chue Hong, Neil and Haines, Robert and Baker, James and Bliven, Spencer and Willighagen, Egon and Pérez-Suárez, David and Konovalov, Alexander},
  year         = 2021,
  month        = {8},
  doi          = {10.5281/ZENODO.5171937},
  url          = {https://zenodo.org/record/5171937},
  note         = {Publisher: Zenodo Version Number: 1.2.0},
  abstractnote = {CITATION.cff files are plain text files with human- and machine-readable citation information for software. Code developers can include them in their repositories to let others know how to correctly cite their software. This is the specification for the Citation File Format.}
}

@article{FennerBlogCff,
  title        = {A step forward for software citation: GitHub’s enhanced software citation support},
  author       = {Fenner, Martin},
  year         = 2021,
  month        = {8},
  journal      = {Front Matter},
  doi          = {10.53731/r9531p1-97aq74v-ag78v},
  url          = {https://blog.front-matter.io/posts/step-forward-for-software-citation},
  abstractnote = {On August 19, GitHub announced software citation support in GitHub repositories. Citation information provided by users (using a CITATION.cff YAML file in the root directory of the default branch) is parsed and made available as bibtex file or formatted citation, ...}
}

@misc{CffSphinxExample,
    author = {Breitbach, Gisbert and Geyer, Beate and Kleeberg, Ulrike and Meyer, Elke and Onken, Reiner and Sommer, Philipp S.},
    title = {Binding Regulations for Storing Data as netCDF Files},
    url = {https://gitlab.hzdr.de/hcdc/hereon-netcdf/hereon-netcdf-en}
}

@inproceedings{RSD-2018,
	address = {Amsterdam, Netherlands},
	title = {Painting the picture of software impact with the {Research} {Software} {Directory}},
	isbn = {978-1-5386-9156-4},
	url = {https://ieeexplore.ieee.org/abstract/document/8588632},
	doi = {10.1109/eScience.2018.00013},
	booktitle = {2018 {IEEE} 14th {International} {Conference} on e-{Science} (e-{Science})},
	publisher = {IEEE},
	author = {Spaaks, Jurriaan H and Klaver, Tom and Verhoeven, Stefan and Maassen, Jason and Bakker, Tom and van der Ploeg, Atze and van Werkhoven, Ben and van Hage, Willem and van Nieuwpoort, Rob V},
	year = {2018},
	pages = {23--24},
}

@inproceedings{SWHArchive,
  title        = {Archiving and Referencing Source Code with Software Heritage},
  author       = {Di Cosmo, Roberto},
  year         = {2020},
  booktitle    = {Mathematical Software – ICMS 2020},
  publisher    = {Springer International Publishing},
  series       = {Lecture Notes in Computer Science},
  pages        = {362–373},
  doi          = {10.1007/978-3-030-52200-1_36},
  isbn         = {978-3-030-52200-1},
  place        = {Cham},
  abstractnote = {Software, and software source code in particular, is widely used in modern research. It must be properly archived, referenced, described and cited in order to build a stable and long lasting corpus of scientific knowledge. In this article we show how the Software Heritage universal source code archive provides a means to fully address the first two concerns, by archiving seamlessly all publicly available software source code, and by providing intrinsic persistent identifiers that allow to reference it at various granularities in a way that is at the same time convenient and effective.We call upon the research community to adopt widely this approach.},
  editor       = {Bigatti, Anna Maria and Carette, Jacques and Davenport, James H. and Joswig, Michael and de Wolff, Timo},
  collection   = {Lecture Notes in Computer Science}
}

@misc{Zenodo,
  doi = {10.25495/7GXK-RD71},
  url = {https://www.zenodo.org/},
  author = {{European Organization For Nuclear Research} and {OpenAIRE}},
  language = {en},
  title = {Zenodo: Research. Shared.},
  publisher = {CERN},
  year = {2013},
  abstractnote = {Zenodo is a general purpose repository that enables researchers, scientists, projects and institutions to share, preserve and showcase multidisciplinary research results (data, software and publications) that are not part of the existing institutional or subject-based repositories of the research communities. It is founded in the trustworthy CERN data centre.}
}

@article{SARA,
  title        = {SARA-Dienst: Software langfristig verfügbar machen},
  author       = {Rapp, Franziska and Kombrink, Stefan and Kushnarenko, Volodymyr and Fratz, Matthias and Scharon, Daniel},
  year         = {2018},
  month        = {7},
  journal      = {o-bib : Das offene Bibliotheksjournal},
  pages        = {92--105},
  doi          = {10.5282/O-BIB/2018H2S92-105},
  url          = {https://www.o-bib.de/article/view/2018H2S92-105},
  note         = {Artwork Size: 92-105 Seiten Publisher: o-bib : Das offene Bibliotheksjournal / Herausgeber VDB},
  abstractnote = {Software spielt in vielen Disziplinen eine wichtige Rolle im Forschungsprozess. Sie ist entweder selbst Gegenstand der Forschung oder wird als Hilfsmittel zur Erfassung, Verarbeitung und Analyse von Forschungsdaten eingesetzt. Zur Nachvollziehbarkeit der durchgeführten Forschung sollte Software langfristig verfügbar gemacht werden. Im SARA-Projekt zwischen der Universität Konstanz und der Universität Ulm wird ein Dienst entwickelt, der versucht die Einschränkungen bereits bestehender Angebote aufzuheben. Dies beinhaltet u.a. die Möglichkeit, die gesamte Entwicklungshistorie auf einfache Weise mitzuveröffentlichen und für Dritte zur Online-Exploration anzubieten. Zudem bestimmen die Forschenden den Zeitpunkt und Umfang der zu archivierenden/veröffentlichenden Software-Artefakte selbst. Der SARA-Dienst sieht auch die Möglichkeit vor, eine Archivierung ohne Veröffentlichung vorzunehmen. Der geplante Dienst verbindet bereits bestehende Publikations- und Forschungsinfrastrukturen miteinander. Er ermöglicht aus der Arbeitsumgebung der Forschenden heraus eine Archivierung und Veröffentlichung von Software und unterstützt Forschende dabei, bereits prozessbegleitend Zwischenstände ihrer Forschung festzuhalten. Aufgrund seines modularen Aufbaus kann der SARA-Dienst in unterschiedlichen Szenarien zum Einsatz kommen, beispielsweise als kooperativer Dienst für mehrere Einrichtungen. Er stellt eine sinnvolle Ergänzung zu bestehenden Angeboten im Forschungsdatenmanagement dar. Software plays an important role in many scientific disciplines. Whether software itself is the research focus or whether software tools are used to create, process and analyse data – software should be available for the long term to make the research process reproducible. In the context of the SARA project conducted by the University of Konstanz and Ulm University, a service is being developed which aims to avoid restrictions of existing services. This includes the possibility to easily publish the whole change history and make it available for others to explore online. Additionally, the researchers decide when and what they want to archive/publish. The SARA service also allows for archiving of software without making it publicly accessible. It connects existing publication and research infrastructures. Researchers can trigger a publication of software from their research environment and are encouraged to publish software artefacts already during the research process. The new service can be used in various scenarios due to its modular design, for example as a cooperative service for several institutions. The SARA Service is a useful addition to already existing research data management services.}
}

@software{sw:codemeta-generator,
  title        = {Codemeta Generator},
  author       = {Software Heritage},
  license      = {AGPL-3.0},
  year         = {2019},
  month        = {10},
  url          = {https://codemeta.github.io/codemeta-generator},
  repository   = {https://github.com/codemeta/codemeta-generator},
  abstractnote = {A (client-side) web application to generate CodeMeta documents (aka codemeta.json).}
}

@software{sw:cffconvert,
  title        = {cffconvert},
  author       = {Spaaks, Jurriaan H. and Klaver, Tom and Verhoeven, Stefan and Druskat, Stephan and Leoncio, Waldir, Netto},
  year         = {2021},
  month        = {9},
  license      = {Apache-2.0},
  doi          = {10.5281/zenodo.1162057},
  url          = {https://pypi.org/project/cffconvert/},
  repository   = {https://github.com/citation-file-format/cff-converter-python}
}

@software{sw:cffinit,
  title        = {cffinit},
  author       = {Spaaks, Jurriaan H. and Verhoeven, Stefan and Diblen, Faruk and Druskat, Stephan and Soares Siqueira, Abel and Garcia Gonzalez, Jesus},
  license      = {Apache-2.0},
  year         = {2021},
  month        = {11},
  doi          = {10.5281/zenodo.1404735},
  url          = {https://citation-file-format.github.io/cff-initializer-javascript},
  repository   = {https://github.com/citation-file-format/cff-initializer-javascript}
}

@softwaremodule{sw:codemeta-crosswalks,
  crossref     = {CodeMetaSchema},
  subtitle     = {Crosswalks},
  author       = {CodeMeta Authors},
  year         = {2020},
  url          = {https://codemeta.github.io/crosswalk/},
  repository   = {https://github.com/codemeta/codemeta/tree/master/crosswalks}
}

@software{sw:biblatex-software,
  title        = {biblatex-software},
  author       = {Di Cosmo, Roberto},
  year         = {2020},
  month        = {6},
  license      = {LPPL-1.3c},
  url          = {https://ctan.org/pkg/biblatex-software},
  repository   = {https://gitlab.inria.fr/gt-sw-citation/bibtex-sw-entry}
}

@software{sw:swh-save-action,
  title        = {swh-save-action},
  author       = {Druskat, Stephan},
  year         = {2021},
  license      = {MIT},
  url          = {https://github.com/marketplace/actions/save-to-software-heritage},
  repository   = {https://github.com/sdruskat/swh-save-action},
  abstractnote = {A GitHub Action that saves a GitHub repository to the Software Heritage Archive}
}

@software{sw:dataverse-uploader-action,
  title        = {Dataverse Uploader GitHub Action},
  author       = {Trisovic, Ana and Durbin, Philip},
  year         = {2021},
  license      = {MIT},
  url          = {https://github.com/marketplace/actions/dataverse-uploader-action},
  repository   = {https://github.com/IQSS/dataverse-uploader},
  abstractnote = {GitHub Action to publish repository content on Dataverse }
}

@software{sw:SoMEF,
  title        = {Software Metadata Extraction Framework (SoMEF)},
  author       = {Garijo, Daniel et al.},
  year         = 2019,
  month        = {10},
  license      = {MIT},
  doi          = {10.5281/zenodo.3477929},
  url          = {https://somef.readthedocs.io},
  repository   = {https://github.com/KnowledgeCaptureAndDiscovery/somef},
  abstractnote = {A command line interface for automatically extracting relevant information from readme files.}
}

@softwareversion{sw:codemeta2cff,
  title        = {CodeMeta2CFF},
  author       = {Morrell, Thomas E.},
  year         = 2021,
  month        = {7},
  publisher    = {CaltechDATA},
  license      = {BSD-3-Clause},
  institution  = {Caltech},
  doi          = {10.22002/D1.2048},
  url          = {http://dx.doi.org/10.22002/D1.2048},
  repository   = {https://github.com/caltechlibrary/codemeta2cff},
  abstractnote = {GitHub Action converting a codemeta file to CITATION.cff}
}

@softwareversion{sw:AMES,
  title        = {AMES},
  author       = {Morrell, Thomas E. and Doiel, Robert},
  year         = {2022},
  month        = {1},
  publisher    = {CaltechDATA},
  license      = {BSD-3-Clause},
  institution  = {Caltech},
  doi          = {10.22002/D1.8964},
  url          = {http://dx.doi.org/10.22002/D1.8964},
  repository   = {https://github.com/caltechlibrary/ames},
  abstractnote = {Manage metadata from different sources. The examples in the package are specific to Caltech repositories, but could be generalized. This package is currently in development and will have additional sources and matchers added over time.}
}

@software{sw:rsd,
    author = {Spaaks, Jurriaan H. and Klaver, Tom and Verhoeven, Stefan and Diblen, Faruk and Maassen, Jason and Tjong Kim Sang, Erik and Pawar, Pushpanjali and Meijer, Christiaan and Ridder, Lars and Kulik, Lode and Bakker, Tom and van Hees, Vincent and Bogaardt, Laurens and Mendrik, Adriënne and van Es, Bram and Attema, Jisk and van Hage, Willem and Ranguelova, Elena and van Nieuwpoort, Rob and Gey, Ronny and Zach, Hoskins},
    doi = {10.5281/zenodo.1154130},
    license = {Apache-2.0},
    month = {12},
    title = {Research Software Directory},
    url = {https://research-software.nl},
    repository = {https://github.com/research-software-directory/research-software-directory},
    year = {2020}
}

@softwaremodule{sw:swh-deposit-cli,
    title = {Software Heritage},
    subtitle = {Deposit},
    author = {Software Heritage},
    year = {2017},
    institution = {Inria},
    license = {GPL-3},
    url = {https://docs.softwareheritage.org/devel/swh-deposit/index.html},
    repository= {https://forge.softwareheritage.org/source/swh-deposit},
}

@software{sw:rodare-bridge,
    title = {Invenio-GitLab},
    author = {Tobias Huste},
    year = {2019},
    institution = {HZDR},
    license = {GPL-3},
    url = {https://rodare.hzdr.de},
    repository = {https://gitlab.hzdr.de/rodare/invenio-gitlab}
}

@software{sw:invenio-github,
    title = {Invenio-GitHub},
    author = {Invenio Software},
    year = {2021},
    institution = {CERN},
    license = {GPL-2.0},
    url = {https://invenio-github.readthedocs.io},
    repository = {https://github.com/inveniosoftware/invenio-github}
}

@article{sw:opencarp,
    title = {The {openCARP} simulation environment for cardiac electrophysiology},
    volume = {208},
    issn = {0169-2607},
    url = {https://www.sciencedirect.com/science/article/pii/S0169260721002972},
    doi = {https://doi.org/10.1016/j.cmpb.2021.106223},
    journal = {Computer Methods and Programs in Biomedicine},
    author = {Plank, Gernot and Loewe, Axel and Neic, Aurel and Augustin, Christoph and Huang, Yung-Lin and Gsell, Matthias A. F. and Karabelas, Elias and Nothstein, Mark and Prassl, Anton J. and Sánchez, Jorge and Seemann, Gunnar and Vigmond, Edward J.},
    year = {2021},
    pages = {106223},
}

@softwaremodule{sw:opencarp-ci,
    crossref = { sw:opencarp },
    subtitle = { CI },
    institution = { KIT },
    license = { Apache-2.0 },
    url = {https://opencarp.org/},
    repository = { https://git.opencarp.org/openCARP/openCARP-CI }
}

@software{sw:zenodraft,
    title        = {zenodraft},
    author       = {Spaaks, Jurriaan H.},
    year         = 2021,
    month        = {7},
    publisher    = {Zenodo},
    license      = {Apache-2.0},
    doi          = {10.5281/ZENODO.5046392},
    url          = {https://zenodo.org/record/5046392},
    repository   = {https://github.com/zenodraft/zenodraft},
    abstractnote = {CLI to manage depositions on Zenodo or Zenodo Sandbox.}
}

@softwaremodule{sw:zenodraft-action,
    crossref     = {sw:zenodraft},
    subtitle     = {Zenodraft GitHub Action},
    url          = {https://github.com/marketplace/actions/zenodraft},
    repository   = {https://github.com/zenodraft/action},
    abstractnote = {GitHub Action to automate drafting depositions on Zenodo or Zenodo Sandbox.}
}

@article{jaySoftwareMustBe2021,
  title = {Software {{Must}} Be {{Recognised}} as an {{Important Output}} of {{Scholarly Research}}},
  author = {Jay, Caroline and Haines, Robert and Katz, Daniel S.},
  year = {2021},
  month = apr,
  journal = {International Journal of Digital Curation},
  volume = {16},
  number = {1},
  pages = {6},
  issn = {1746-8256},
  doi = {10.2218/ijdc.v16i1.745},
  copyright = {Copyright (c) 2021 Caroline Jay, Robert Haines, Daniel S. Katz},
  langid = {english}
}

\end{document}